\documentclass[aps,showpacs,twocolumn,eqsecnum,superscriptaddress]{revtex4}

\newcommand{\bea}{\begin{eqnarray}}
\newcommand{\eea}{\end{eqnarray}}
\newcommand{\beq}{\begin{equation}}
\newcommand{\eeq}{\end{equation}}

\usepackage{latexsym}
\usepackage{amssymb}
\usepackage{amsmath}
\usepackage{epsfig}

\begin{document}


\title{Constraint and gauge shocks in one-dimensional numerical relativity}


\author{Bernd Reimann}
\affiliation{Instituto de Ciencias Nucleares, Universidad Nacional
Aut{\'o}noma de M{\'e}xico, A.P. 70-543, M{\'e}xico D.F. 04510, M{\'e}xico}
\affiliation{Max Planck Institut f\"ur Gravitationsphysik,
Albert Einstein Institut, Am M\"uhlenberg 1, 14476 Golm, Germany}

\author{Miguel Alcubierre}
\affiliation{Instituto de Ciencias Nucleares, Universidad Nacional
Aut{\'o}noma de M{\'e}xico, A.P. 70-543, M{\'e}xico D.F. 04510, M{\'e}xico}

\author{Jos\'e A. Gonz\'alez}
\affiliation{Institute for Gravitational Physics and Geometry.
Penn State University, University Park, PA 16802}

\author{Dar\'\i o N\'u\~nez}
\affiliation{Instituto de Ciencias Nucleares, Universidad Nacional
Aut{\'o}noma de M{\'e}xico, A.P. 70-543, M{\'e}xico D.F. 04510, M{\'e}xico}


\date{March 3, 2005}


\begin{abstract}

We study how different types of blow-ups can occur in systems of
hyperbolic evolution equations of the type found in general
relativity.  In particular, we discuss two independent criteria that
can be used to determine when such blow-ups can be expected.  One
criteria is related with the so-called geometric blow-up leading to
gradient catastrophes, while the other is based upon the ODE-mechanism
leading to blow-ups within finite time.  We show how both mechanisms
work in the case of a simple one-dimensional wave equation with a
dynamic wave speed and sources, and later explore how those blow-ups
can appear in one-dimensional numerical relativity.  In the latter
case we recover the well known ``gauge shocks'' associated with
Bona-Masso type slicing conditions.  However, a crucial result of this
study has been the identification of a second family of blow-ups
associated with the way in which the constraints have been used to
construct a hyperbolic formulation.  We call these blow-ups
``constraint shocks'' and show that they are formulation specific, and
that choices can be made to eliminate them or at least make them less
severe.

\end{abstract}


\pacs{
04.20.Ex,   
04.25.Dm,   
95.30.Sf    
\qquad Preprint numbers: AEI-2004-110, IGPG-04/11-5
}


\maketitle


\section{Introduction}
\label{sec:intro}

When studying the Cauchy problem of field theories in physics, one has
to worry about the existence and uniqueness of solutions to the system
of evolution equations considered.  In mathematical terms, one is
interested in working with problems that are well-posed, by which one
understands that a unique solution exists (at least locally), and
solutions are stable in the sense that small changes in the initial
data produce small changes in the solution.  In this respect, one
usually looks for either symmetric or strongly hyperbolic systems of
equations since Cauchy problems for such systems are known to be well
posed under very general conditions~\cite{Kreiss89}.

In the case of general relativity, the Cauchy problem was studied since
the 50's with the pioneering work of Choquet-Bruhat~\cite{Choquet52},
and by the mid 80's a number of hyperbolic reductions were known~(see
for example~\cite{Choquet62,Choquet80,Choquet83,Friedrich85}, and more
recent reviews in~\cite{Friedrich96,Friedrich:2000qv}).  Still, those
reductions played a minor role in numerical relativity, where
practically all work using the Cauchy approach was based on the
Arnowitt-Deser-Misner (ADM) system of
equations~\cite{Arnowitt62,York79}.  Interest in hyperbolic
formulations in numerical relativity started in the early 90's, with
the work of Bona and Masso~\cite{Bona89,Bona92,Bona94b}, but continued
as a small side branch for a number of years.  This situation remained
until Baumgarte and Shapiro showed in~\cite{Baumgarte99} that a
reformulation of the ADM equations originally proposed by Nakamura,
Oohara and Kojima~\cite{Nakamura87}, and Shibata and
Nakamura~\cite{Shibata95}, had far superior numerical stability
properties than ADM. Baumgarte and Shapiro attributed this to the fact
that the new formulation, which has since become known as BSSN, had a
``more hyperbolic flavor''.  This rather informal statement was later
put on firmer ground in~\cite{Alcubierre99e,Sarbach02a,Beyer:2004sv},
and today it is understood that ADM is only weakly hyperbolic (and
thus not well posed)~\cite{Frittelli:2000uj}, whereas BSSN is strongly
hyperbolic~\cite{Sarbach02a,Beyer:2004sv}.

The recognition by the numerical relativity community of the fact that
well-posedness is a crucial ingredient for having long term stable and
well behaved numerical simulations~(see \cite{Gustafsson95}) has led,
in recent years, to an explosion in the search for ever more general
hyperbolic reductions of the Einstein evolution equations.  At this
time many such hyperbolic formulations exist, several of which have
dozens of free parameters~(see for
example~\cite{Kidder01a,Sarbach:2002gr}).  The large number of ways in
which one can construct strongly or even symmetric hyperbolic
formulations has taken us to a situation where there are now many more
proposed formulations than numerical groups capable of testing them.
At the same time, there is a growing realization that in some respects
well-posedness is not enough, as empirically some hyperbolic
formulations have proven to be far more robust than others.  Some work
has been done on the analytic side trying to understand what makes
some hyperbolic formulations better suited for numerical work.  In
particular one can mention the work of Shinkai and
Yoneda~\cite{Shinkai:2001nd}, where the propagation of constraints for
different formulations is studied by linearizing the evolution system
around the Schwarzschild background and looking for the eigenvalues of
the evolution matrix in Fourier space, and the work of Lindblom and
Scheel~\cite{Lindblom:2002et}, where the rate of growth of the
constraint violation is analyzed for a family of symmetric hyperbolic
formulations using the fact that for such systems one can construct an
``energy norm''.  The consensus is that one should look beyond the
principal part of the system, and study the effect of the source terms
on the stability.

In this paper, we want to focus on a different aspect that can
differentiate between hyperbolic formulations and that has been so far
overlooked.  Well-posed formulations are known to have well behaved
solutions locally, but there is no guarantee that these solutions can
exist beyond a certain finite time. In fact, on physical grounds we
expect solutions to fail after a finite time in some circumstances,
due for example to the formation of singularities in gravitational
collapse. But there is another way in which solutions can become
singular after a finite time, the best example of which is the
formation of shock waves in hydrodynamics.  In general relativity, and
particularly in vacuum, we do not expect these type of shock waves to
develop.  Nevertheless, one must remember that the evolution equations
evolve more than just the physical degrees of freedom.  In particular,
there are also gauge degrees of freedom that can cause coordinate
singularities to arise during the evolution. In~\cite{Alcubierre97a}
one of us showed that coordinate singularities caused by the crossing
of the characteristic lines associated with the propagation of the
gauge can in fact easily form.  These so-called ``gauge shocks'' are
now known to be responsible for the fact that some gauge choices are
much better behaved than others.  In particular, in
reference~\cite{Alcubierre02b} it was shown that the well known
``1+log'' slicing condition, which empirically has been found to be
very robust, is in fact the only member of its family that avoids
gauge shocks approximately.  More recently it has been found that, for
the evolution of Brill wave spacetimes that are very close to the
critical threshold for black hole formation, the use of shock avoiding
slicing conditions is crucial~\footnote{P. Diener, private
communication.}.

But there are degrees of freedom other than the physical and gauge
modes that also appear in the evolution equations of general
relativity.  These extra degrees of freedom have to do with the
violation of the constraints, and even though for physical initial
data they should vanish identically, truncation errors make their
presence unavoidable in numerical simulations.  It is therefore very
important to understand how these constraint violating modes behave
analytically.  A main result of this paper is the recognition that
constraint modes can also give rise to the development of blow-ups in
a finite time that are very similar to the gauge shocks studied
before.  These blow-ups are a property of the specific form of the
evolution equations and their effects can be significantly reduced if
one chooses carefully how the constraints are used when constructing a
hyperbolic system.  In this paper we show how blow-ups can arise in
spherically symmetric relativity, and how they can best be avoided by
modifying the evolution system.  We believe that the study of such
constraint shocks might help us understand why otherwise well-posed
and ``nice'' formulations might behave poorly in numerical
simulations.

A comment about our terminology is in order. Borrowing the language
from hydro-dynamics, throughout the paper we will refer in a somewhat
loose way to blow-ups as ``shocks'', though this term strictly only
refers to blow-ups caused by the crossing of characteristic lines.

This paper is organized as follows.  In
Section~\ref{sec:hyperbolicity} we introduce the concept of
hyperbolicity and describe two different criteria that can be used to
determine when blow-ups in the solutions to hyperbolic systems of
equations can be expected.  Section~\ref{sec:wave} introduces the
simple one-dimensional wave equation with sources and a dynamic wave
speed as an example of how these blow-ups develop.  In
Section~\ref{sec:Einstein} we apply the blow-up criteria to the
evolution equations of general relativity.  We start with the simple
case of 1+1 dimensions, where we recover the well known gauge shocks,
and later study the case of spherical symmetry where we find that
constraint shocks can also arise. We conclude in
Section~\ref{sec:conclusions}.


\section{Hyperbolicity and shocks}
\label{sec:hyperbolicity}

The system of evolution equations we are interested in analyzing are
the evolution equations for the Cauchy problem of general relativity.
In particular we are interested in studying the appearance of singular
non-physical solutions.  Such an analysis can be best made using the
characteristic structure of hyperbolic systems, so we will start from
the definition of hyperbolicity.  We will also concentrate on systems
with only one spatial dimension as this makes the analysis so much
simpler.  The important point of what happens in the multi-dimensional
case is a matter for future research.  Notice that one-dimensional
systems are in fact relevant in general relativity, as they can represent 
the evolution of systems with, for example, spherical symmetry.

There is one important point that should be mentioned.  Throughout this
section, and in the rest of the paper, we manipulate differential
equations by assuming that partial derivatives in time and space
commute.  This type of manipulation leaves smooth (``classical'')
solutions unchanged, but can easily change the speed of propagation of
shock waves~\cite{Leveque92}.  Still, in this paper we will only be
interested in smooth solutions, and we will consider the development
of a shock as a pathology.  Our whole emphasis is in finding ways to
avoid shocks.


\subsection{Hyperbolic systems}
\label{sec:hyperbolic}

We will consider quasi-linear systems of evolution equations that can
be split into two subsystems with the following structure
\begin{eqnarray}
\partial_t \vec{u} &=& - {\bf M}^1_u \; \partial_x \vec{u}
- {\bf M}^2_u \; \vec{K} \; , \\
\partial_t \vec{K} &+& {\bf M}^1_K \; \partial^2_x \vec{u}
+ {\bf M}^2_K \; \partial_x \vec{K} 
= \vec{p}_K(\vec{u},\partial_x \vec{u},\vec{K}) \; . \qquad
\end{eqnarray}
Here $\vec{u}$ and $\vec{K}$ are $n$ and $m$ dimensional vectors
respectively, and $\bold{M}^{1,2}_u$ and $\bold{M}^{1,2}_K$ are
matrices whose coefficients may depend on the $u$'s, but not on the
$K$'s.

In order to have a first order system we will introduce the spatial
derivatives \mbox{$D_i := \partial_x u_i$} as extra independent
variables, whose evolution equations are obtained directly from those
of the $u$'s, 
\begin{equation}
\partial_t \vec{D} + \partial_x \left( {\bf M}^1_u \; \vec{D} 
+ {\bf M}^2_u \; \vec{K} \right) = 0 \; .
\label{eq:DPDE}
\end{equation}

In the following we will always assume that the initial data satisfies
the constraints \mbox{$D_i = \partial_x u_i$}.  This implies that
spatial derivatives of the $u$'s can always be substituted for $D$'s
and treated as source terms.

Let us now define the $(n+m)$ dimensional vector \mbox{$\vec{v}:=
  (\vec{D}, \vec{K})$}. We can then rewrite the system of evolution
equations as
\begin{eqnarray}
\partial_t \vec{u} &=& - {\bf M}(\vec{u}) \; \vec{v} \; , \label{eq:uPDE} \\
\partial_t \vec{v} &+& {\bf A}(\vec{u}) \; \partial_x \vec{v}
= \vec{q}_v(\vec{u},\vec{v}) \; ,
\label{eq:vPDEmatrix}
\end{eqnarray}
where ${\bf M}$ and ${\bf A}$ are $n \times (n+m)$ and
\mbox{$(n+m) \times (n+m)$} matrices, 
\begin{equation}
{\bf M} = \left( \begin{array}{cc}
{\bf M}^1_u & {\bf M}^2_u
\end{array} \right) \; ,
\qquad
{\bf A} = \left( \begin{array}{cc}
{\bf M}^1_u & {\bf M}^2_u \\
{\bf M}^1_K & {\bf M}^2_K \rule{0mm}{5mm}
\end{array} \right) \; ,
\end{equation}
and where the source vector $\vec{q}_v$ is given by
\begin{equation}
\vec{q}_v = \left(- \sum_{i=1}^n 
D_i \left[ (\partial_{u_i} {\bf M}^1_u) \vec{D}
+ (\partial_{u_i} {\bf M}^2_u) \vec{K} \right] , \vec{p}_K \right) .
\end{equation}

In our primary example, the Einstein equations, the vector $\vec{u}$
consists of gauge variables and components of the \hbox{3-metric},
whereas $\vec{v}$ contains both variables associated with the spatial
derivatives of the gauge variables and metric components (the $D$'s)
and also variables arising from the time derivatives of the metric
components (the $K$'s).  Note, furthermore, that the source terms
$\vec{q}_v$ appearing on the right-hand side of~(\ref{eq:vPDEmatrix})
are in general functions of both the $u$'s and $v$'s (typically
quadratic on the $v$'s).

The system of equations above will be hyperbolic if the matrix
$\bf{A}$ has \mbox{$(n+m)$} real eigenvalues $\lambda_i$.  Furthermore, it will
be strongly hyperbolic if it has a complete set of eigenvectors
$\vec{\xi}_i$,
\begin{equation}
{\bf A} \; \vec{\xi}_i = \lambda_i \vec{\xi}_i \; .
\label{eq:eigenvectorsR}
\end{equation}
If we denote the matrix of column eigenvectors by ${\bf R}$,
\begin{equation}
{\bf R} = \left( \vec{\xi}_1 \cdots \vec{\xi}_{(n+m)} \right) \; , 
\end{equation}
then the matrix $\bf{A}$ can be diagonalized as
\begin{equation}
{\bf R}^{-1} {\bf A R} = \begin{rm}{\textbf{diag}}\end{rm}
\left[ \lambda_1, \cdots, \lambda_{(n+m)} \right]
= {\bf \Lambda} \; .
\label{eq:diagonalize}
\end{equation}
Notice that for systems with only one spatial dimension the otherwise
important distinction between strongly and symmetric hyperbolic
systems does not arise.

For a strongly hyperbolic system we define the eigenfields as
\begin{equation}
\label{eq:w}
\vec{w} = {\bf R}^{-1} \vec{v} \; .
\end{equation}
By analyzing the time evolution of the eigenfields we now want to
study by which mechanisms blow-ups (i.e. singularities) in the
solutions can arise.  As pointed out in~\cite{Alinhac}, there are
basically two different blow-up mechanisms which are, somewhat
misleadingly, referred to as ``geometric blow-up'' and the
``ODE-mechanism''.  Since in the first case the derivative of an
evolution variable, and in the second case an evolution variable
itself, becomes infinite within a finite time, the names ``gradient
catastrophe''~\cite{John86} and ``blow-up within finite time'' are
probably more appropriate. In the following sections we will explain
the basic idea behind these two mechanisms using as a prototype a
simple scalar equation, and we will show how both mechanisms are in
fact closely related.  We will also point out how these mechanisms
generalize to systems of PDE's.


\subsection{Geometric blow-up and linear degeneracy}
\label{sec:geometric}


\subsubsection{Scalar conservation laws}
\label{sec:scalar}

The first mechanism responsible for blow-ups involves only
quasi-linear systems of equations.  Here the solution $u$ under
consideration has a well-defined limit at a given point and only the
derivatives of $u$ become infinite there.  Typical examples of this
situation are obtained when solving scalar conservation equations of
the form
\begin{equation}
\partial_t u + \partial_x F(u)
= \partial_t u + a(u) \; \partial_x u = 0 \; ,
\label{eq:PDEscalar1D}
\end{equation}
where $a(u) := \partial F(u) / \partial u$. The blow-up is then due to
the focusing of characteristics at a point, and the mechanism is
referred to as ``geometric blow-up''.  Taking the simplest nonlinear
function $F(u) = u^2/2$, we obtain Burgers' equation
\begin{equation}
\partial_t u + u \; \partial_x u = 0 \; ,
\label{eq:Burgers}
\end{equation}
which is frequently discussed in the literature as an example of a
genuinely nonlinear PDE leading to shock formation (see for
example~\cite{Hopf50,John86}).  The solution of Eq.~(\ref{eq:Burgers})
can be interpreted as a time-dependent one-dimensional velocity field
$u$.  The equation then states that the characteristics (i.e. the
``flow lines'') have zero acceleration, that is $du/dt = \partial_t u
+ (dx/dt) \; \partial_x u =\partial_t u + u \; \partial_x u = 0$,
which means that particles following those trajectories move with
constant velocity $u = dx/dt$.  However, unless the initial velocity
distribution $u_0(x)$ is a non-decreasing function of $x$ (so that the
particles ``spread out''), eventually a particle with higher velocity
will collide with one ahead of it having a lower velocity.  In
particular, as particles initially at rest are not moving at all, $u$
is forced to become singular at a finite time if its initial velocity
distribution has compact support (except in the trivial case when
$u_0(x)$ vanishes everywhere).


\subsubsection{Linear degeneracy}
\label{sec:lineardegeneracy}

The previous argument coming from Burgers' equation~(\ref{eq:Burgers})
can be easily generalized to the case of Eq.~(\ref{eq:PDEscalar1D}).
Consider two locations $x_1$ and $x_2$ with $x_1 < x_2$ and
corresponding initial values $u_1 = u_0(x_1)$ and $u_2 = u_0(x_2)$.
Just as before, the values of $u$ are conserved along characteristics,
so unless $a(u(x))$ is purely increasing in $x$, it is always possible
to find locations such that $a(u_1) > a(u_2)$, so the characteristic
lines on the left go faster than those on the right (as $a(u_i)$
represents their velocity).  One may readily verify that the lines
will intersect at a time given by
\begin{equation}
t^* = - \left( \frac{x_1 - x_2}{a(u_1) - a(u_2)} \right) \; .
\end{equation}
At the point of intersection $u$ has to take both values $u_1$ and
$u_2$, so a unique solution ceases to exist.  When this happens the
spatial derivative of $u$ becomes infinite and the differential
equation breaks down, in other words no smooth solution of
(\ref{eq:PDEscalar1D}) exists after $t=t^*$.

For smooth initial data, taking the limit \mbox{$|x_1 - x_2| \to 0$}
in the expression for $t^*$, we see that this gradient catastrophe will occur
at a finite time
\begin{equation}
t^* = - \frac{1}{\begin{rm}{min}\end{rm} \left[ \partial_x
a(u_0)\right]} = - \frac{1}{\begin{rm}{min}\end{rm} \left[ a'(u_0)
\partial_x u_0 \right]}
\end{equation}
in the ``future'' if initially we have \mbox{$\partial_x a(u_0(x)) < 0$}, 
whereas the problem arises in the ``past'' if \mbox{$\partial_x a(u_0(x)) > 0$}
 holds initially.  Since in general one can not guarantee that every initial 
data set one would like to use satisfies such a condition, the criteria for 
not forming a shock demands that the function $a(u)$ should be linear, 
that is $a'(u) = 0$.

The above argument can also be generalized to (strongly) hyperbolic
systems of equations, see~\cite{Lax73}.  For such systems, the
condition one needs in order to avoid the formation of shocks
associated with the propagation of a given eigenfield $w_i$ is for
this eigenfield to be ``linearly degenerate'', which means that the
eigenvalue $\lambda_i$ associated with $w_i$ must be constant along
integral curves of the corresponding eigenvector $\vec{\xi}_i$:
\begin{equation}
\nabla_v \lambda_i \cdot \vec{\xi}_i
= \frac{\partial \lambda_i}{\partial w_i}
= \sum_{j=1}^{(n+m)} \frac{\partial \lambda_i}{\partial v_j}
\frac{\partial v_j}{\partial w_i} = 0 \; .
\label{eq:LD}
\end{equation}


\subsection{ODE-mechanism and the source criteria}
\label{sec:ODEmechanism}


\subsubsection{ODE's with quadratic sources}
\label{sec:ODEsources}

In ODE's, PDE's and systems of PDE's, an evolution variable itself can
become infinite at a point by a process of ``self-increase'' in the
domain of influence leading to this point.  In a somewhat misleading
way, the underlying mechanism has been given the name
``ODE-mechanism'' (see~\cite{Alinhac}), since prototype examples are
based on simple ODE's such as
\begin{equation}
\label{eq:ODEexample}
\frac{du}{dt} = c u^2 \; , \qquad c = {\rm constant} \neq 0 \; .
\end{equation}
For non-trivial initial data the solution of~(\ref{eq:ODEexample}) is
\begin{equation}
u(t) = \frac{u_0}{1 - u_0 c t} \; , \qquad u_0 \neq 0 \; .
\end{equation}
This solution clearly blows up at a finite time
given by
\begin{equation}
t^* = \frac{1}{u_0 \, c} \; ,
\end{equation}
either in the past or in the future, depending on the sign of $u_0 c$.
One can also expect such blow-ups to happen in the case when $c$ is not a 
constant but instead a function of time.  If $c(t)$ is bounded, one can 
apply theorems 1 and 2 of~\cite{Lax64}: 
Supposing that the function $c(t)$ satisfies the inequality $0 < C < c(t)$ 
for $0 \le t \le T$, and that $u_0$ is positive, then $T < 1/(u_0 C)$
since $u(t)$ for all positive $t$ is bounded from below by $u_0/(1 -
u_0 C t)$.  Similarly, supposing that $c(t)$ satisfies the inequality
$| c(t) | < \tilde{C}$, then the initial value problem has a solution
for at least \mbox{$|t| < 1/| u_0 \tilde{C} |$}.


\subsubsection{The source criteria for avoiding shocks} 
\label{sec:sourcecriteria}

Let us now go back to our original system of equations
(\ref{eq:uPDE})-(\ref{eq:vPDEmatrix}).  Multiplying
Eq.~(\ref{eq:vPDEmatrix}) from the left by ${\bf R}^{-1}$ we find
\begin{eqnarray}
\partial_t \left( {\bf R}^{-1} \vec{v} \right)
+ \left( {\bf R}^{-1} {\bf A} {\bf R} \right)
\partial_x \left( {\bf R}^{-1} \vec{v} \right) \hspace{25mm} \nonumber \\
= {\bf R}^{-1} \vec{q}_v + \left[ \partial_t {\bf R}^{-1}
+ \left( {\bf R}^{-1} {\bf A} {\bf R} \right) \;
\partial_x  {\bf R}^{-1} \right] \vec{v} \; , \hspace{5mm}
\end{eqnarray}
which, by making use of (\ref{eq:diagonalize}) and (\ref{eq:w}),
yields
\begin{equation}
\label{eq:wPDE}
\partial_t \vec{w} + {\bf \Lambda} \partial_x \vec{w}
= \vec{q}_w \; ,
\end{equation}
where
\begin{equation}
\vec{q}_w := {\bf R}^{-1} \vec{q}_v + \left[ \partial_t  {\bf R}^{-1}
+ {\bf \Lambda} \partial_x  {\bf R}^{-1} \right] \vec{v} \; .
\label{eq:qw}
\end{equation}

In this way we have obtained an evolution system where on the
left-hand side of (\ref{eq:wPDE}) the different eigenfields $w_i$ are
decoupled. However, in general the equations are still coupled through
the source terms $q_{w_i}$.  In particular, if the original sources
were quadratic in the $v$'s, we will have
\begin{equation}
\frac{dw_i}{dt} = \partial_t w_i + \lambda_i \partial_x
w_i = \sum_{j,k=1}^{(n+m)} c_{ijk} w_j w_k + {\cal O}(w) \; ,
\end{equation}
where $d/dt := \partial_t + \lambda_i \partial_x$ is the derivative
along the corresponding characteristic.  As pointed out for a similar
system in~\cite{John74,John90}, here the $c_{iii} w_i^2$ component of
the source term can be expected to dominate, so mixed and lower order
terms can be neglected.  Though we have no proof of this statement in
the general case, one can expect it to be true at least for systems
with distinct eigenspeeds, as mixed terms will then be suppressed when
pulses moving at different speeds separate from each other.  The
effect of the term $c_{iii} w_i^2$, on the other hand, will remain
even as the pulse moves.  The numerical simulations presented in the
following sections show empirical evidence that reinforces this
argument.

We can then rewrite the above equation as
\begin{equation}
\label{eq:dwdt=ww}
\frac{dw_i}{dt} \approx c_{iii} w_i^2 \; ,
\end{equation}
which has precisely the form of the ODE studied in the previous
section.  In order to avoid a blow-up one would then have to demand
that the coefficients $c_{iii}$ vanish.  We call this the ``source
criteria'' for avoiding blow-ups.

There is a very important property of our system of equations
regarding the coefficients $c_{iii}$ that come from the source terms
$\vec{q}_w$.  From~(\ref{eq:qw}) one could expect contributions to these 
coefficients coming both from the original sources $\vec{q}_v$ and 
from the term in brackets involving derivatives of ${\bf R}^{-1}$.  
However, one can show that this is not the case and for the systems 
under study the contributions to $c_{iii}$ coming from the term in 
brackets cancel out, that is, all contributions to $c_{iii}$ come only 
from the original sources $\vec{q}_v$.

In order to see this we start by rewriting the term inside brackets on
the right hand side of~(\ref{eq:qw}) as
\begin{equation}
\partial_t {\bf R}^{-1}
+ {\bf \Lambda} \: \partial_x  {\bf R}^{-1} =
\sum_{l=1}^n \left( \partial_t u_l + {\bf \Lambda} \: \partial_x u_l \right)
\partial_{u_l} {\bf R}^{-1} \; .
\end{equation}
Since both the time and space derivatives of $u_l$ can be written in
terms of $v$'s, and the above term multiplies the vector $\vec{v}$ in
Eq.~(\ref{eq:qw}), it is clear that this term will give rise to
quadratic terms in the $v$'s, and hence in the $w$'s.  The question is
whether these quadratic terms will produce a contribution to the coefficients 
$c_{iii}$.  From the last equation it is clear that no
such contribution will exist if the following condition is satisfied
\begin{equation}
\frac{\partial}{\partial w_i} \left( \partial_t u_l
+ \lambda_i \partial_x u_l \right) = 0 ,
\quad \forall \; i \leq (n+m) , \; l \leq n .
\label{eq:eigencondition}
\end{equation}
We will now show that this condition is indeed always satisfied.
Notice first that, from the definition of the eigenfields $\vec{w}$
and the matrix ${\bf R}$, one can easily see that
\begin{equation}
\frac{\partial}{\partial w_i} = \vec{\xi}_i \cdot \nabla_v \; ,
\end{equation}
with the eigenvector $\vec{\xi}_i$ corresponding to the eigenvalue 
$\lambda_i$.  Now, from equation~(\ref{eq:uPDE}) and the definition of 
the $D$'s we have
\begin{equation}
\partial_t u_l + \lambda_i \: \partial_x u_l
= - \sum_{j=1}^{(n+m)} M_{lj} v_j + \lambda_i D_l \; ,
\end{equation}
which implies that
\begin{equation}
\vec{\xi}_i \cdot \nabla_v \left( \partial_t u_l
+ \lambda_i \: \partial_x u_l \right)
= \lambda_i \: {\xi_i}_l - \sum_{j=1}^{(n+m)} M_{lj} \: {\xi_i}_j \: ,
\end{equation}
where ${\xi_i}_j$ is the $j$ component of the vector $\vec{\xi_i}$,
and where we have used the fact that the first $n$ components of
$\vec{v}$ are precisely the $D$'s (remember that by construction $l
\leq n$).

To finish the proof we now use the fact that the first $n$ rows of the
matrix ${\bf A}$ are given by the matrix ${\bf M}$, and also the fact
that $\vec{\xi}_i$ is an eigenvector of ${\bf A}$ with eigenvalue
$\lambda_i$, which implies
\begin{equation}
\sum_{j=1}^{(n+m)} M_{lj} \: {\xi_i}_j
= \sum_{j=1}^{(n+m)} A_{lj} \: {\xi_i}_j
= \lambda_i \: {\xi_i}_l \; ,
\end{equation}
from which we finally find
\begin{equation}
\vec{\xi}_i \cdot \nabla_v \left( \partial_t u_l
+ \lambda_i \: \partial_x u_l \right) = 0 \; .
\end{equation}
This completes the proof that condition~(\ref{eq:eigencondition})
always holds (as long as the constraints $D_i=\partial_x u_i$ are
satisfied), which in turn means that the term in square brackets
in~(\ref{eq:qw}) does not contribute to the coefficients $c_{iii}$.

We want to make another important comment here: As the eigenvectors
$\vec{\xi}_i$ diagonalizing the matrix ${\bf A}$ are obtained only up
to an arbitrary rescaling, also the eigenfields $w_i$ are not unique.
In particular, any $w_i$ can be multiplied by an arbitrary function of
the $u$'s to obtain $\tilde{w}_i = \Omega_i (\vec{u}) \: w_i$.
However, since the $v$'s are related to derivatives of the $u$'s, such
a rescaling will introduce new quadratic source terms, so one would in
general not expect the coefficients $c_{iii}$ to be invariant under
rescalings of the eigenfields.

Remarkably, for the systems of the type~(\ref{eq:uPDE})-(\ref{eq:vPDEmatrix}) 
that we are interested in, it turns out that such rescalings of the 
eigenfields have no effect on the coefficients $c_{iii}$.  The proof of 
this is again related to condition~(\ref{eq:eigencondition}).  In general, 
if we rescale the eigenfunctions as 
$\tilde{w}_i = \Omega_i (\vec{u}) \; w_i$, we will
find that
\begin{eqnarray}
\partial_t \tilde{w}_i &+& \lambda_i \partial_x \tilde{w}_i =
\Omega \left( \partial_t w_i + \lambda_i \partial_x w_i \right) \nonumber \\
&& \hspace{7mm} + w_i \sum_{l=1}^n \partial_{u_l} \Omega \: \left( \partial_t 
u_l + \lambda_i \partial_x u_l \right) \nonumber \\
&=& \Omega \: q_{w_i} + w_i \sum_{l=1}^n \partial_{u_l} \Omega \:
\left( \partial_t 
u_l + \lambda_i \partial_x u_l \right) . \hspace{5mm}
\end{eqnarray}
From this we see that, although the rescaling does introduce new
quadratic terms, condition~(\ref{eq:eigencondition}) guarantees that
no new contributions to the coefficient $c_{iii}$ will arise, {\em
i.e.} the source criteria for avoiding blow-ups is invariant with
respect to rescalings of the eigenfunctions (again, as long as the
constraints $D_i=\partial_x u_i$ are satisfied).


\subsubsection{Is the source criteria necessary and sufficient
in order to avoid blow-ups?}

A question one might immediately ask is whether the source criteria
introduced above is necessary and sufficient to avoid blow-ups.
Although we have no proof at this time, numerical experiments (such as
those shown in later sections) indicate that whenever the source
criteria is not satisfied blow-ups do develop, which would support our
conjecture that the criteria is indeed necessary in order to avoid
blow-ups.

About it being sufficient, it clearly is not.  This can be seen from
the following example.  Consider the following system of two equations,
\begin{eqnarray}
\partial_t v_1 + \partial_x v_2 &=& v_1^2 + v_2^2 \; , \\
\partial_t v_2 + \partial_x v_1 &=& - 2 v_1 v_2 \; ,
\end{eqnarray}
which can easily be diagonalized to find
\begin{eqnarray}
\partial_t w_+ + \partial_x w_+ &=& w_-^2 \; , \\
\partial_t w_- - \partial_x w_- &=& w_+^2 \; ,
\end{eqnarray}
with $w_\pm := v_1 \pm v_2$.  The diagonalized system clearly
satisfies the source criteria.  Now, take initial data such that
$v_1=k= {\rm const}$ and $v_2=0$ in a large spatial region.  This
implies that in that region $w_1=w_2=k$.  Since the spatial
derivatives vanish and both fields are in fact equal, the equations
above are of the form (\ref{eq:ODEexample}) and the fields will blow
up in finite time (provided the region where the fields where
initially equal is large enough).  This initial data is clearly very
special, but it does show that the source criteria is not sufficient
in order to avoid a blow-up.  Nevertheless, for more generic data this
situation will be very rare.

We have in fact performed numerical experiments with systems of the
above form, but generalizing the source terms to
\begin{equation}
\partial_t w_\pm \pm \partial_x w_\pm
= a_\pm w^2_\pm + b_\pm w_+ w_- + c_\pm w^2_\mp \; .
\end{equation} 
When using Gaussians with a small amplitude of order ${\cal O}(\epsilon)$ as 
initial data for $w_\pm$, then whenever the coefficients $a_\pm$ are non-zero 
one typically finds that blow-ups occur on a timescale of order 
${\cal O}(1/\epsilon)$.  
If, on the other hand, $a_\pm=0$ and one has only mixed terms and/or terms
quadratic in the other eigenfield in the sources, blow-ups again
eventually develop, but now on a timescale of order ${\cal
O}(1/\epsilon^2)$.  Hence, if one is interested in propagating small
perturbations, then satisfying the source criteria should allow one to
obtain longer evolutions.


\subsection{Relationship between the different blow-up mechanisms}

In order to understand the relationship between the geometric blow-up
and the ODE-mechanism, we will study a system of two variables
constructed from the simple scalar conservation
law~(\ref{eq:PDEscalar1D}) by introducing either the time or space
derivative of the function $u$ as an extra independent variable.

We start by introducing $D := \partial_x u$ as a new variable.  One
then obtains the system
\begin{eqnarray}
\label{eq:DuxA}
\partial_t u &=& - a(u) \; D  \; , \\
\label{eq:DuxB}
\partial_t D &+& a(u) \; \partial_x D = - a'(u) \; D^2 \; ,
\end{eqnarray}
where the evolution equation for $D$ has been found by differentiating
(\ref{eq:PDEscalar1D}) with respect to $x$ and exchanging the order of
$\partial_t$ and $\partial_x$.

As we are interested in studying solutions of the original scalar
conservation law, but seen from a different perspective, will only
consider initial data such that the constraint $D := \partial_x u$ is
satisfied. Remembering that along a characteristic line
of~(\ref{eq:PDEscalar1D}) $u$ is constant (and hence also $a(u)$ and
$a'(u)$), the equation
\begin{equation}
\label{eq:dDdt=DD}
\frac{dD}{dt} = \partial_t D + a(u) \; \partial_x D = -
a'(u) \; D^2 \; ,
\end{equation}
arising from~(\ref{eq:DuxB}) can be easily integrated.  We find that,
along the characteristic, the following relation holds
\begin{equation}
D(t) = \frac{D_0}{1 + D_0 a'(u) t} \; .
\end{equation}
This clearly becomes infinite at a time $t^*$ given by
\begin{equation}
\label{eq:blowupDux}
t^* = - \frac{1}{D_0 a'(u)} \; .
\end{equation}

Let us now introduce $K := \partial_t u$ instead of $D$ as an extra
variable. We then obtain the system
\begin{eqnarray}
\label{eq:KutA}
\partial_t u &=& K \; , \\
\label{eq:KutB}
\partial_t K &+& a(u) \; \partial_x K = \frac{a'(u)}{a(u)} \; K^2 \; .  
\end{eqnarray}
Here the evolution equation for $K$ has been derived by taking a
partial derivative with respect to $t$ of~(\ref{eq:PDEscalar1D}).
As before, by integrating the equation
\begin{equation}
\label{eq:dKdt=KK}
\frac{dK}{dt} = \partial_t K + a(u) \partial_x K
= \frac{a'(u)}{a(u)} \; K^2 \; ,
\end{equation}
along the characteristic one finds
\begin{equation}
K(t) = \frac{K_0}{1 - (K_0 a'(u) \; / a(u)) \; t} \; ,
\end{equation}
which diverges at a time given by
\begin{equation}
\label{eq:blowupKut}
t^* = \frac{a(u)}{K_0 a'(u)} \; .
\end{equation}

These two examples are nothing more than our original scalar
equation~(\ref{eq:PDEscalar1D}) in disguise.  However, they are in
fact linearly degenerate by the definition given above as the only
eigenvalue $a(u)$ is independent of $D$ and $K$, respectively.  They
still give rise to a blow-up, as they should, but this time the
blow-up appears through the ODE-mechanism instead of the original
geometric blow-up mechanism.  Notice that, from~(\ref{eq:blowupDux})
and~(\ref{eq:blowupKut}), one can conclude that a condition for not
having a blow-up in finite time is $a'(u) = 0$, which is the same
condition we found in Sec.~\ref{sec:lineardegeneracy} above.  This
shows clearly that what can be considered a geometric blow-up of a
given variable $u$ can always be reinterpreted as an ODE-type blow-up
of its derivatives, so both blow-up mechanisms are closely related.


\subsection{Indirect Linear Degeneracy}
\label{sec:indirect}

Linear degeneracy turns out to be insufficient for avoiding blow-ups
in the particular case of system (\ref{eq:uPDE})-(\ref{eq:vPDEmatrix})
for two reasons.  The first reason has to do with the presence of
non-vanishing source terms $\vec{q}_v$ and has been discussed in
Sec.~\ref{sec:ODEmechanism} above.  The other reason is simply the
fact that for these type of systems the eigenvalues of the
characteristic matrix ${\bf A}$ depend only the $u$'s and not on the
$v$'s, which means that all eigenfields are linearly degenerate in a
trivial way. We have already seen an example of this in the previous
section when we considered the simple scalar conservation law and
introduced derivatives as extra independent variables.

For this reason the concept of ``indirect linear degeneracy'' was
introduced in~\cite{Alcubierre97a}.  This simply replaces the
eigenvalue $\lambda_i$ in equation~(\ref{eq:LD}) by its time
derivative:
\begin{equation}
\nabla_v \dot{\lambda}_i \cdot \vec{\xi}_i
= \frac{\partial \dot{\lambda}_i}{\partial w_i}
= \sum_{j=1}^{(n+m)} \frac{\partial \dot{\lambda}_i}{\partial v_j}
\frac{\partial v_j}{\partial w_i} = 0 \; .
\label{eq:iLD}
\end{equation}
This new condition yields non-trivial results for the system
(\ref{eq:uPDE})-(\ref{eq:vPDEmatrix}) if the time derivatives of the
$u$'s, appearing when differentiating $\lambda_i$ with respect to
time, depend on the corresponding $w_i$.

It is in fact not difficult to see where the indirect linear
degeneracy condition comes from.  Consider the system of two equations
\begin{eqnarray}
\partial_t u &=& p(u,v,\partial_x u) \; , \\
\partial_t v &+& \lambda(u) \, \partial_x v = q(u,v,\partial_x u) \; ,
\end{eqnarray}
with $p$ linear in $v$ and $\partial_x u$. We now extend the above
system by introducing the variable $D:= \partial_x u$.  This means
that the sources $p$ and $q$ are now functions of $(u,v,D)$.  The full
system will then be
\begin{eqnarray}
\partial_t u &=& p \; , \\
\partial_t D &-& \partial_x p = 0 \; , \\ 
\partial_t v &+& \lambda(u) \, \partial_x v = q \; ,
\end{eqnarray}
which is exactly of the form~(\ref{eq:uPDE})-(\ref{eq:vPDEmatrix}).
Let us for a moment assume that $q=0$.  In that case it is clear that
$v$ will be constant along the characteristics lines $x = x_0 +
\lambda(u) \; t$.  The simplest example is obtained when
\mbox{$\lambda(u) = u$} and $p = v - u D$, since then we find that
along the characteristics \mbox{$du/dt = v$} (provided that the
constraint $D=\partial_x u$ remains satisfied).  This means that along
those lines we have $u=u_0 + v t$, so the characteristics have
constant acceleration given by $v$ (since $u$ is the characteristic
speed).  If initially $v_0(x)=v(t=0,x)$ has negative slope in a given
region, the characteristics are then guaranteed to cross (as lines
behind accelerate faster than those in front). At the point where this
happens the gradient of $v$ will become infinite and we will have a
blow-up.  For cases when $p$ is a different function one can not
integrate the equations exactly, but the same general idea will
hold. Of course, when the source term $q$ is not zero one could
imagine that $q$ can be chosen in such a way as to avoid the crossing
of characteristics, but such a choice would clearly not be generic.
The only way to be sure that there will be no blow-up is to ask for
$\partial p / \partial v = 0$.  Indirect linear degeneracy is simply
the generalization of this condition to the case of a system with more
equations.

The argument given above, however, is clearly not rigorous.  Indirect
linear degeneracy is therefore still a more or less {\em ad hoc}
condition.  Part of the reason for discussing it here is precisely to study
its relevance in different cases by numerical experiments.  As
our results in the following sections show, indirect linear degeneracy
and the source criteria often yield the same conditions for avoiding
blow-ups.  When they do not, the source criteria seems to be more
important.  Exploring the link between indirect linear degeneracy and
the source criteria is something that should be further investigated,
and we are currently working in that direction.


\section{The wave equation with sources and a dynamic wave speed}
\label{sec:wave}


\subsection{Blow-up formation}
\label{sec:wavetheory}

As an example for the type of evolution systems studied in the
previous sections we will consider the simple scalar wave equation
with sources,
\begin{equation}
\partial^2_{t} u - c^2(u) \: \partial^2_{x} u 
= q(u,\partial_t u, \partial_x u) \; .
\label{eq:wavePDE}
\end{equation}
Here we allow the wave speed $c$ to be a function of the wave function
$u$.  The source term $q$, on the other hand, can depend both on $u$
and its first derivatives.  Introducing $D = \partial_x u$ and $K =
\partial_t u$, we can rewrite the wave equation as
\begin{eqnarray}
\partial_t u &=& K \; , \\
\partial_t D &-& \partial_x K = 0 \; , \\
\partial_t K &-& c^2 \partial_x D = q \; , 
\end{eqnarray}  
which is of the form (\ref{eq:uPDE})-(\ref{eq:vPDEmatrix}).  One can
readily verify that the eigenvalues of the characteristic matrix,
\begin{equation}
{\bf A} = \left( \begin{array}{cc}
0 & -1 \\
-c^2 & 0 \end{array} \right) \; ,
\end{equation}
are $\lambda_\pm = \pm c$, with corresponding eigenfields
(the normalization is chosen for convenience)
\begin{equation}
w_{\pm} = \frac{1}{2} \left( K \mp c D \right) \; .
\end{equation}

The linear degeneracy criteria then is trivially satisfied since the
eigenvalues depend only on $u$.  However, when we calculate the time
derivative of the eigenvalues we find
\begin{equation}
\dot{\lambda}_\pm = \pm c' \partial_t u = \pm c' K =
\pm c' \left( w_{+} + w_{-} \right) \; .
\end{equation}
The indirect linear degeneracy condition~(\ref{eq:iLD}) asks for
the derivatives $\partial \dot{\lambda}_\pm / \partial w_\pm$
to vanish, which implies
\begin{equation}
c' = 0 \; .
\end{equation}
This is no longer trivial and corresponds to what one would in fact
expect: The wave speed should be independent of the wave function
if we want no shocks to develop.

Let us now turn to the source criteria. We find
\begin{equation}
\frac{dw_\pm}{dt} = \frac{1}{2} \; q(u,w_\pm) + \frac{c'}{c} \left(
w_{+} w_{-} - w_{\mp}^2 \right) \; .
\label{eq:wave_wdot}
\end{equation}
Notice that the evolution equation for a given eigenfield $w_\pm$
contains no quadratic terms on itself other than those that might come
from $q$ (it has only mixed terms and terms quadratic in the other
eigenfield).  The source criteria then demands that the source term
$q$ should be free of the quadratic terms $w_{+}^2$ and $w_{-}^2$.  If
we assume that $q$ is of the form
\begin{eqnarray}
q &=& {\cal A} \; c^2 D^2 + {\cal B} \; c \; D K + {\cal C} K^2
\nonumber \\
&=& 
\left( {\cal A} - {\cal B} + {\cal C} \right) w_{+}^2 
-2 \left( {\cal A} - {\cal C} \right) w_{+} w_{-} \nonumber \\
&+& \left( {\cal A} + {\cal B} + {\cal C} \right) w_{-}^2 \; ,
\end{eqnarray}
with ${\cal A}$, ${\cal B}$ and ${\cal C}$ arbitrary functions of $u$,
it follows that in order to avoid blow-ups these functions have to
satisfy
\begin{equation}
{\cal B} = 0 \; , \qquad {\cal A} + {\cal C} = 0 \; .
\label{eq:waveSC}
\end{equation}


\subsection{Numerical results}
\label{sec:wavenumerics}

We have performed a series of numerical simulations for wave equations
which satisfy or violate indirect linear degeneracy and/or the source
criteria.  The results from these simulations are summarized in
Table~\ref{tab:waveequation}.  All simulations have been performed
using a method of lines with fourth order Runge-Kutta integration in
time, and standard second order centered differences in space (with no
artificial dissipation added~\footnote{Artificial dissipation (or some
other more advanced technique like shock capturing) is crucial to
follow shock waves, but here we don't want to go beyond shock
formation, so we don't actually need it.  In hydrodynamics there are
physical principles that allow one to follow the solution past the
shock formation (we know that physically there is no true
discontinuity), but in this case there is no analog of that.}).

As initial data we have taken $u(t=0)=1$ and hence $D(t=0)=0$,
together with the derivative of a Gaussian for the time derivative of
$u$, i.e.
\begin{equation}
K(t=0) = - \left( 2 \kappa x /\sigma^2 \right)
\exp \left( -x^2/\sigma^2 \right) \; .
\end{equation}
Here we used the derivative of a Gaussian and not a simple
Gaussian in order to excite perturbations where $u$ is both
smaller and larger than its initial value.

For all the runs shown here we have used the particular values
$\kappa=0.1$ and $\sigma=0.3$.  These rather strong and localized
perturbations are motivated by the fact that we wanted to see shock
formation early, in particular before the variable $u$ changes sign
(since the first two examples with eigenspeeds given by $\pm u$ are not 
strongly hyperbolic for $u = 0$).  However, also for other 
values of $\kappa$ and $\sigma$ we have seen qualitatively very similar 
behavior (and whether or not $u$ crosses zero does not seem to play a 
role in the formation of shocks).

For the runs with highest resolution we have used $80,000$
grid points and a resolution of $\Delta x = 5 \times 10^{-4}$, which
places the boundaries at $\pm 20$, together with time steps of $\Delta
t = \Delta x/2$.  In addition, for each evolution variable we have
computed the convergence factor $\eta$ which, using three runs with
high ($u^h$), medium ($u^m$) and low ($u^l$) resolutions differing in
each case by a factor of two, can be calculated as e.g.
\begin{equation}
\eta_u = \frac{\frac{1}{N_i} \sum_{i = 1}^{N_i} | u^m_i - u^l_i |}
{\frac{1}{N_j} \sum_{j = 1}^{N_j} | u^h_j - u^m_j |} \; .
\end{equation}

In the plots we show four different convergence factors: We denote
with a triangle the convergence factors obtained when comparing runs
with $80,000$, $40,000$ and $20,000$ grid points and a spatial
resolution of $5 \times 10^{-4}$, $10^{-3}$ and $2 \times 10^{-3}$.
We use boxes, diamonds and stars to denote the convergence factors
when gradually lowering all three resolutions by a factor of two.  For
second order convergence we expect $\eta \simeq 4$.

\begin{table}
\begin{tabular}{|c|c|c|c|}
\hline
i.l.d. & s.c. & $(c,q)$ & result \\
\hline
no  & no  & $c=u$, $q=2uD^2$ & blow-up
(Fig.~\ref{fig:no-no}) \\
no  & yes & $c=u$, $q=2(uD^2 - K^2/u)$ & blow-up?
(Fig.~\ref{fig:no-yes}) \\
yes & no  & $c=1$, $q=2D^2$ & blow-up
(Fig.~\ref{fig:yes-no}) \\
yes & yes & $c = 1$, $q = 2( D^2 - K^2)$ & no blow-up
(Fig.~\ref{fig:yes-yes}) \\
\hline
\end{tabular}
\caption{Summary of results from simulations of a wave equation that
satisfies (yes) or violates (no) both indirect linear degeneracy
(i.l.d.) and the shock criteria (s.c.).}
\label{tab:waveequation}
\end{table}

\begin{figure}
\epsfxsize=85mm
\epsfysize=100mm
\epsfbox{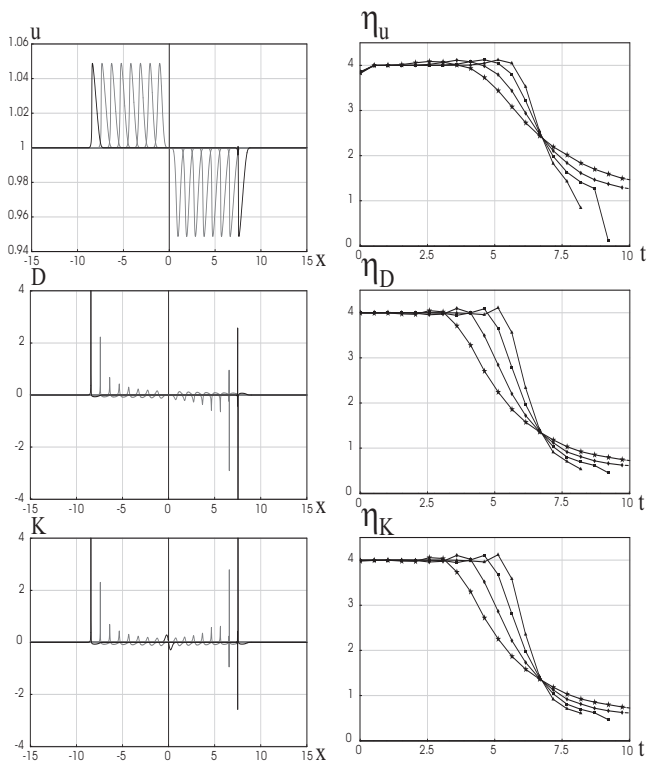}
\caption{Results for the case when both criteria are violated.  This
  simulation corresponds to the time derivative of Burgers' equation,
  for which we have \mbox{$c=u$} and \mbox{$q=2uD^2$}.  In the left
  panel a sharp gradient is clearly visible on $u$, and large peaks
  can also be seen on $D$ and $K$.  In the right panel we see a clear
  loss of convergence that becomes sharper as resolution is increased
  in the order ``star, diamond, box and triangle'', indicating a
  blow-up at a time \mbox{$t \simeq 7$}.}
\label{fig:no-no}
\end{figure}

The first test we have done corresponds to a case that violates both
blow-up criteria.  We obtain such a system by simply taking a time
derivative of Burgers' equation.  We find $\partial^2_{t} u - u^2
\partial^2_{x} u = 2 u (\partial_x u)^2$ and hence identify
\begin{equation}
c = u \; , \qquad q = 2 u D^2 \; .
\end{equation}
Results for this simulation can be found in Figure~\ref{fig:no-no}.
In the left panel we show snapshots of the evolution of the variables
$u$, $D$ and $K$ in steps of $\Delta t = 1$, and in the right panel we
show convergence factors for the previously mentioned series of
different resolutions.  From the figure we clearly see that, as
expected, shocks do form, with large gradients developing on $u$ and
large peaks on $D$ and $K$.  Moreover, from the convergence plots we
see that there is a clear loss of convergence, and as the resolution
is increased, this loss of convergence becomes more sharply centered
around a specific time $t \simeq 7$, indicating that the blow-up
happens at this time.

\begin{figure}
\epsfxsize=85mm
\epsfysize=100mm
\epsfbox{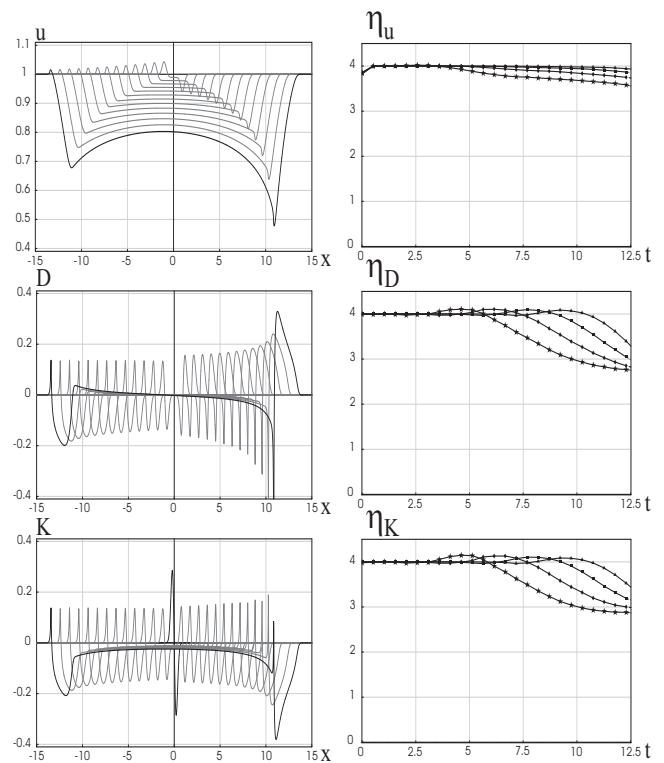}
\caption{Simulation for a case that violates indirect linear
  degeneracy, but satisfies the source criteria (\mbox{$c=u$},
  \mbox{$q=2(uD^2 - K^2/u)$}).  The left panel shows that $D$ and $K$
  develop large peaks, while $u$ develops a sharp gradient but no
  peak. The convergence plots on the right panel indicate also loss of
  convergence in $D$ and $K$, however convergence seems to improve
  with resolution.}
\label{fig:no-yes}        
\end{figure}

For the second example we have chosen a situation where indirect linear
degeneracy is violated but the source criteria holds.  Numerical results
for the case
\begin{equation}
c = u \; , \qquad q = 2(uD^2 - K^2/u) 
\end{equation}
are shown in Figure~\ref{fig:no-yes} (a simpler case arising for a
vanishing source term, $q = 0$, yields very similar results).  The
figure shows large peaks developing in both $D$ and $K$, and a sharp
gradient developing in $u$.  The convergence plots show some loss of
convergence for the lower resolutions but, in contrast with the
previous example, convergence seems to improve with resolution.  This
would seem to indicate that although sharp gradients do develop, a
real blow-up has not occurred.  Nevertheless, such sharp gradients are
difficult to resolve numerically, so their presence is undesirable.

\begin{figure}
\epsfxsize=85mm
\epsfysize=100mm
\epsfbox{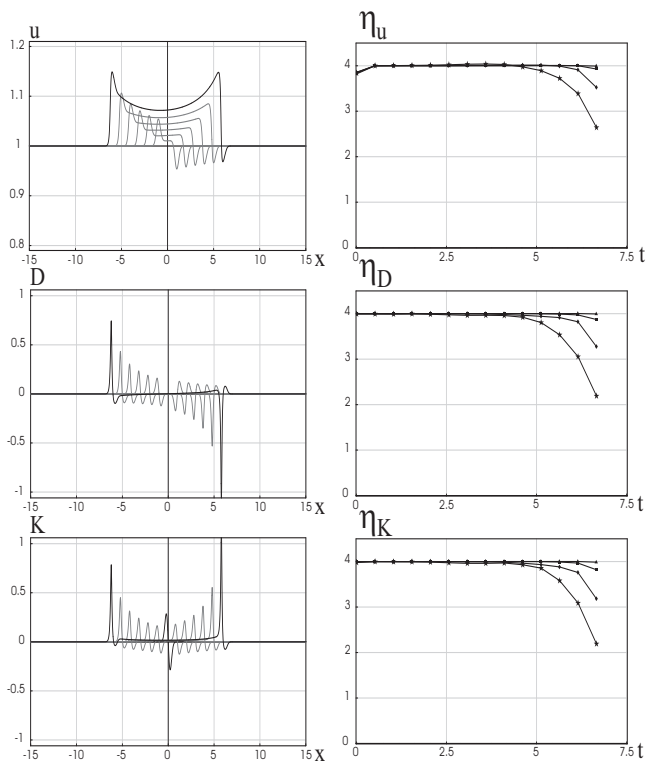}
\caption{Results of a simulation for the case that satisfies indirect
  linear degeneracy, but violates the source criteria (\mbox{$c=1$},
  \mbox{$q=2D^2$}). The left panel shows that $D$ and $K$ develop
  large peaks, while $u$ develops both a peak and a large gradient.
  The right panel shows that convergence is lost at low resolutions
  but improves at higher resolutions, until a final time of $t \simeq
  7$ when runs at all resolutions crash, indicating a true blow-up.}
\label{fig:yes-no}        
\end{figure}

The third simulation corresponds to a case that satisfies indirect
linear degeneracy, but violates the source criteria, with wave speed
and source term given by
\begin{equation}
c = 1 \; , \qquad q = 2 D^2 \; .
\end{equation}
Results for this run are shown in Figure~\ref{fig:yes-no}.  As before,
we see that both $D$ and $K$ are developing large peaks. The evolution
variable $u$ is developing both a large peak and a large gradient.
The convergence plots show a loss of convergence at lower resolutions
that improves as the resolution is increased.  However in this case
all runs crash at $t \simeq 7$, indicating that a blow-up has indeed
occurred at that time.

\begin{figure}
\epsfxsize=85mm
\epsfysize=100mm
\epsfbox{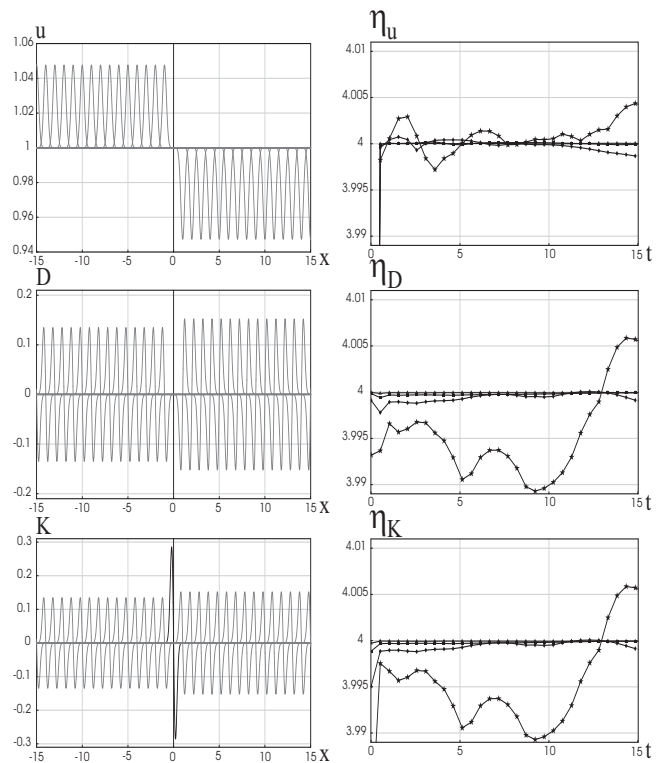}
\caption{Simulation for the case that satisfies both blow-up avoiding
  criteria.  In this case we have taken \mbox{$c=1$} and
  \mbox{$q=2(D^2 - K^2)$}, and we find wave-like behavior with no
  evidence of a blow-up.}
\label{fig:yes-yes}
\end{figure}

Finally, our last test corresponds to the case when both criteria are
satisfied, with the wave speed $c$ and source term $q$ given by
\begin{equation}
c = 1 \; , \qquad q = 2( D^2 - K^2) \; .
\label{eq:yes-yes}
\end{equation}
Results for this simulation are shown in Figure~\ref{fig:yes-yes}.  We
see that the solution behaves in a wave-like manner, with no evidence
of a blow-up.  This result is reinforced by the convergence plots
indicating that we have close to second order convergence during the
whole run for all resolutions considered, with no evidence of loss of
convergence at any time (notice the change in scale with respect to
previous plots).

From the previous simulations it is clear that, for the scalar wave
equation with a dynamic wave speed and sources, sharp gradients and
blow-ups are only avoided when both indirect linear degeneracy and the
source criteria are satisfied.  In particular, the case with $c=1$ and
$q=2(D^2 - K^2)$ behaves very similar to what one would expect from
the standard wave equation with unit wave speed and a vanishing source
term.

This observation can be easily understood by generalizing an example
suggested by L. Nirenberg in \cite{Klainerman80}: Smooth solutions
that extend globally in time will always exist if one can find a
smooth transformation of the form $\tilde{u} = \tilde{u}(u)$, for
which $\tilde{u}$ satisfies the standard wave equation,
$\partial^2_{t} \tilde{u} - \partial^2_{x} \tilde{u} = 0$, see
\cite{John81,Strauss89}.  One may readily verify that for our
particular example with wave speed and source term given by
(\ref{eq:yes-yes}), this is indeed the case for the variable
$\tilde{u} = \exp (2u)$.


\section{The Einstein equations}
\label{sec:Einstein}

In the previous sections we have described how blow-ups can be
produced in systems of hyperbolic equations, and what conditions need
to be satisfied in order for these to be avoided.  We have also
considered one simple example, the wave equation with sources and a
dynamic wave speed.  We will now turn our attention to the system we
are most interested in, namely the evolution equations of general
relativity.  In this paper, we will restrict ourselves to two cases,
``toy'' 1+1 relativity, and spherically symmetric relativity, and
leave the important three-dimensional case for a future work.

We believe that it is important to mention here the main results which
will be presented in this section.  In the first place, we will
recover the results regarding ``gauge shocks'' discussed already
in~\cite{Alcubierre97a,Alcubierre97b,Alcubierre02b}.  But the most
important result that we will show is the fact that, for the
spherically symmetric case, one can also identify a second family of
blow-ups that are not associated with the gauge but rather with the
violation of the constraints.  We will refer to such blow-ups as
``constraint shocks'', since they are clearly associated with the way
in which the constraints have been added to the evolution 
equations.  These constraint shocks will correspond to blow-ups in the
hamiltonian and momentum constraints at a finite time as the numerical 
example at the end of Sec.~\ref{sec:Minkwoski_constraint} 
shows~\footnote{A question arises as to whether one could predict 
the ``constraint shocks'' by applying the source criteria directly to the 
constraint evolution system.  The answer is that one can not, since in the 
constraint evolution system the sources are linear in the constraints, so 
if we keep all coefficients constant no blow-up would be expected (at worst 
one would have exponential growth).  But this is inconsistent, as in fact the 
coefficients are not constant and depend on the metric and its derivatives.}.

Since the blow-up analysis assumes that we have a strongly hyperbolic
system, we will in each case begin by constructing such a hyperbolic
system for the Einstein equations.  Notice that there is no unique way
to obtain hyperbolic evolution systems from the Einstein equations and
we will use this fact to explicitly construct formulations that avoid
constraint shocks.


\subsection{Einstein equations in 1+1 dimensions}
\label{subsec:1p1}

Let us assume that we have standard general relativity in one spatial
dimension. It is well known that in such a case the gravitational
field is trivial and there are no true dynamics.  However, one can
still have nontrivial gauge dynamics that can be used as a simple
example of the type of behavior one can expect in the higher
dimensional case.  We will start from the ``standard''
Arnowitt-Deser-Misner (ADM) equations for one spatial
dimension~\cite{Arnowitt62}, where by standard we mean the version of
York~\cite{York79}.  Since we want a hyperbolic system of evolution
equations that includes the gauge, we will use the Bona-Masso family
of slicing conditions~\cite{Bona94b}:
\begin{equation}
\partial_t \alpha = - \alpha^2 f(\alpha) {\rm tr} K \; ,
\label{eq:BonaMasso}
\end{equation}
where $f=f(\alpha)>0$ identifies the member of the Bona-Masso family
being used (for example, $f=1$ corresponds to harmonic slicing, and
$f=2/\alpha$ to the so-called 1+log slicing). For simplicity we will
also restrict ourselves to the case of vanishing shift vector.

Following \cite{Alcubierre97a}, the two-dimensional vector $\vec{u}$
will consist of the lapse function $\alpha$ and the spatial metric
function $g := g_{xx}$ as components.  The vector $\vec{v}$,
on the other hand, is a three-dimensional vector with components
given by the logarithmic spatial derivatives of $\alpha$ and $g$,
and the unique component of the extrinsic curvature (with mixed
indices).  That is,
\begin{equation}
\vec{u} = \left( \alpha , g \right) \; , \qquad
\vec{v} = \left( D_\alpha , D_g , K \right) \; , 
\end{equation}
where
\begin{equation}
D_\alpha := \partial_x \ln{\alpha} \; , \quad
D_g := \partial_x \ln{g} \: , \quad
K := K_x^x \; .
\end{equation}
(Note that in~\cite{Alcubierre97a} the variable $D = \partial_x g/2$
is used instead of $D_g$ and $K_{xx}$ is used instead of $K^x_x$.)
The fact that we define the $D$'s as logarithmic derivatives instead
of simple derivatives is in order to simplify the resulting equations,
and makes no significant difference in the analysis of
Sec.~\ref{sec:hyperbolicity}.

The ADM evolution equations for the
vectors $\vec{u}$ and $\vec{v}$ turn out to be
\begin{eqnarray}
\partial_t \alpha &=& - \alpha^2 f K \; ,
\label{eq:alphadot} \\
\partial_t g &=& - 2 \alpha g K \; ,
\label{eq:gdot}
\end{eqnarray}
and
\begin{eqnarray}
\partial_t D_\alpha  + \partial_x \left( \alpha f K \right) &=& 0 \; ,
\label{weq:Dadot} \\
\partial_t D_g  + \partial_x \left( 2 \alpha K \right) &=& 0 \; ,
\label{eq:Dgdot} \\
\partial_t K + \partial_x\left( \alpha D_\alpha /g \right)
&=& \alpha \left( K^2 - D_\alpha D_g / 2g \right) \; .
\label{eq:Kdot}
\end{eqnarray}
This system has again the form (\ref{eq:uPDE})-(\ref{eq:vPDEmatrix}).
In particular, the last three equations can be written as
\mbox{$\partial_t \vec{v} + {\bf A} (\vec{u}) \partial_x \vec{v} =
  \vec{q}_v$}, where
\begin{equation}
{\bf A} = \left( \begin{array}{ccc}
0 & 0 & \alpha f \\
0 & 0 & 2\alpha \\
\alpha/g & 0 & 0
\end{array} \right) \; ,
\end{equation}
and
\begin{equation}
\vec{q}_v = \left( \begin{array}{c}
- \alpha \left( f + \alpha f' \right)  D_\alpha K \\
- 2 \alpha D_\alpha K \\
\alpha \left( K^2 - D_\alpha^2/g + D_\alpha D_g / 2 g \right)
\end{array} \right) \; .
\end{equation}

When studying the characteristic structure of the system of equations
above we find the following eigenvalues
\begin{equation}
\lambda_0 = 0 \; , \qquad
\lambda^f_\pm = \pm \alpha \sqrt{\frac{f}{g}} \; , 
\end{equation}
with corresponding eigenfunctions
\begin{eqnarray}
w_0 &=& D_\alpha - \frac{f}{2} D_g \; ,
\label{eq:w0} \\
w^f_\pm &=& \sqrt{f g} K \pm D_\alpha \; .
\label{eq:wfpm}
\end{eqnarray}

The system is therefore strongly hyperbolic as long as $f>0$, with one
eigenfield propagating along the time lines and the other two
propagating with the ``gauge speeds'' $\lambda^f_{\pm} = \pm \alpha
\sqrt{f/g}$.

In order to study the possible formation of shocks for the propagating
eigenfields one can immediately see that the direct linear degeneracy
criteria as formulated in~(\ref{eq:LD}) can not be used, since
$\lambda^f_{\pm}$ does not depend on either $D_\alpha$, $D_g$ or $K$.
The indirect linear degeneracy condition, however, yields
\begin{equation}
\frac{\partial \dot{\lambda}^f_\pm}{\partial w^f_\pm}
= \sum_{j=1}^{(n+m)} \frac{\partial \dot{\lambda}^f_\pm}{\partial v_j}
\frac{\partial v_j}{\partial w^f_\pm}
= \pm \frac{\alpha^2}{2g} \left( 1 - f - \frac{\alpha f'}{2}
\right) = 0 \; ,
\end{equation}  
where the last step comes from expressing the time derivatives of
$\alpha$ and $g$ contained in $\dot{\lambda}^f_{\pm}$ in terms of $K$
using (\ref{eq:alphadot}) and (\ref{eq:gdot}).

For the source criteria, on the other hand, we need to determine the
term quadratic in $w^f_{\pm}$ in the source terms associated to the
evolution equation for $w^f_{\pm}$ itself. We find
\begin{eqnarray}
\frac{dw^f_\pm}{dt} & = & \frac{\alpha}{2 \sqrt{fg}} \; 
\left( 1 - f - \frac{\alpha f'}{2} \right) w^{f \; 2}_{\pm} \nonumber \\
& + & {\cal O} \left( w_0 w^f_{\pm} \; , \; w^f_{\pm} w^f_{\mp} \right) \; .
\end{eqnarray}
Asking for the coefficient of the quadratic term to be zero one finds
\begin{equation}
c_{\pm\pm\pm}^{fff} = \frac{\alpha}{2\sqrt{fg}}
\left( 1 - f - \frac{\alpha f'}{2} \right) = 0 \; .
\end{equation}

It is interesting to note that here both indirect linear degeneracy
and the source criteria yield the same condition for avoiding
blow-ups, namely
\begin{equation}
1 - f - \frac{\alpha f'}{2} = 0 \; .
\label{eq:gaugeshocks}
\end{equation}
The reason why this is so is not completely clear, but it probably
implies that in this case the sources and characteristic speeds are
not independent of each other.

The shock avoiding condition~(\ref{eq:gaugeshocks}) has been studied
before in~\cite{Alcubierre97a,Alcubierre97b,Alcubierre02b}. Its
general solution is
\begin{equation}
f(\alpha) = 1 + \frac{\begin{rm}{const}\end{rm}}{\alpha^2} \; .
\label{eq:shockavoidf}
\end{equation}
Reference~\cite{Alcubierre02b} also considers some approximate
solutions that are more useful for numerical simulations. Notice also
that, since in this simple case we only have gauge dynamics, these
shocks are directly associated with the foliation itself, and for this
reason they are known as ``gauge shocks''.


\subsection{Einstein equations in spherical symmetry}
\label{sec:spherical}


\subsubsection{Standard ADM equations}

In order to generalize the previous system to spherical symmetry, we
start with the spatial line element
\begin{equation} 
dl^{2} = A(t,r) \; dr^{2} + r^2 B(t,r) \; d \Omega^2 \; ,
\end{equation}
where $d \Omega^2 = d\theta^2 + \sin^2 \theta d\phi^2$ denotes the
usual solid angle element.  Note that in this case the vector
$\vec{u}$ will consist of the lapse $\alpha$ and the metric components
$A$ and $B$.  Furthermore, the $v$'s are given by the spatial
derivatives of these quantities
\begin{equation} 
D_\alpha = \partial_r \ln{\alpha} \; , \quad
D_A = \partial_r \ln{A} \; , \quad
D_B = \partial_r \ln{B} \; ,
\end{equation}
together with the extrinsic curvature variables (again with mixed
indices)
\begin{equation}
K_A = K^r_r \; ,\qquad K_B = K^\theta_\theta = K^\phi_\phi \; .
\end{equation}
That is,
\begin{equation}
\vec{u} = \left( \alpha, A, B \right) \; , \quad
\vec{v} = \left( D_\alpha, D_A, D_B, K_A, K_B \right) \; .
\end{equation}

In the following we will assume for simplicity that we are in vacuum
and that the shift vector vanishes. Using the Bona-Masso slicing
condition~(\ref{eq:BonaMasso}), the evolution equations for the $u$'s
become
\begin{eqnarray}
\partial_t \alpha &=& - \alpha^2 f \left( K_A + 2 K_B \right) \; , 
\label{eq:alphadotADM1} \\
\partial_t A &=& - 2 \alpha A K_A \; , 
\label{eq:AdotADM1} \\ 
\partial_t B &=& - 2 \alpha B K_B \; .
\label{eq:BdotADM1} 
\end{eqnarray}

The evolution equation for the $v$'s can be obtained directly from the
ADM equations and the definition of the $D$'s.  These equations can
again be written in the form \mbox{$\partial_t \vec{v} + {\bf A}
  (\vec{u}) \partial_r \vec{v} = \vec{q}_v$}, where the characteristic
matrix is
\begin{equation}
{\bf A} = \left( \begin{array}{ccccc}
0 & 0 & 0 & \alpha f & 2 \alpha f \\
0 & 0 & 0 & 2 \alpha & 0 \\
0 & 0 & 0 & 0 & 2 \alpha \\
\alpha/A & 0 & \alpha/A & 0 & 0 \\
0 & 0 & \alpha/2A & 0 & 0
\end{array} \right) \; ,
\label{eq:matrixADM1}
\end{equation}
and the source terms are given by
\begin{eqnarray}
q_{D_\alpha} &=& - \alpha \left( f + \alpha f' \right)
D_\alpha (K_A + 2 K_B) \; , \\
q_{D_A} &=& - 2 \alpha D_\alpha K_A \; , \\
q_{D_B} &=& - 2 \alpha D_\alpha K_B \; , \\
q_{K_A} &=& - \frac{\alpha}{A} \left[ D_\alpha \left( D_\alpha 
- \frac{D_A}{2} \right) - \frac{D_B}{2} \left( D_A 
- D_B \right) \right. \nonumber \\
&-& \left. A K_A \left( K_A + 2 K_B \right) -
\frac{1}{r} \left( D_A - 2 D_B \right) \right] \; , \\
q_{K_B} &=& - \frac{\alpha}{2 A} \left[ D_B \left( D_\alpha 
- \frac{D_A}{2} + D_B \right) \right. \nonumber \\
&-& 2 A K_B (K_A + 2 K_B) + \frac{1}{r} \left( 2D_\alpha - D_A
+ 4D_B \right) \nonumber \\
&-& \left. \frac{2}{r^2 B} \left( A-B \right) \right] \; .
\end{eqnarray}
Furthermore, the Hamiltonian and momentum constraints take the form
\begin{eqnarray}        
\label{eq:HamCon1}
C_h &:=& - \partial_r D_B
+ \frac{D_B}{2} \left( D_A - \frac{3D_B}{2} \right) \nonumber \\
&+& A K_B \left( 2 K_A + K_B \right) \nonumber \\
&+& \frac{D_A - 3D_B}{r} + \frac{A - B}{r^2 B} = 0  \; , \\
C_m &:=& - \partial_r K_B + \left( \frac{D_B}{2}
+ \frac{1}{r} \right)(K_A - K_B) = 0 \; . \hspace{10mm}
\end{eqnarray}


\subsubsection{ADM equations in new variables}

Rather than working with the standard ADM equations described in the
last section, following~\cite{Alcubierre04a} we will introduce the
``anti-trace'' of the spatial derivatives of the metric components, $D
= D_A - 2D_B$, and the trace of the extrinsic curvature, $K = K_A +
2K_B$, as fundamental variables instead of $D_A$ and $K_A$.  This
choice of variables makes the hyperbolicity analysis more transparent.
The vector $\vec{v}$ will then be
\begin{equation}
\vec{v} = \left( D_\alpha, D, D_B, K, K_B \right) \; ,
\end{equation}
and the evolution of the $u$'s will be given by
\begin{eqnarray}
\partial_t \alpha &=& - \alpha^2 f K \; , 
\label{eq:alphadotADM2} \\
\partial_t A &=& - 2 \alpha A \left( K - 2 K_B \right) \; , 
\label{eq:AdotADM2} \\
\partial_t B &=& - 2 \alpha B K_B \; .
\label{eq:BdotADM2} 
\end{eqnarray}
For the $v$'s, one again obtains a system of the form
\mbox{$\partial_t \vec{v} + {\bf A} (\vec{u}) \partial_r \vec{v} =
\vec{q}_v$}, but this time with characteristic matrix
\begin{equation}
{\bf A} = \left( \begin{array}{ccccc}
0 & 0 & 0 & \alpha f & 0 \\
0 & 0 & 0 & 2 \alpha & -8\alpha \\
0 & 0 & 0 & 0 & 2 \alpha \\
\alpha/A & 0 & 2\alpha/A & 0 & 0 \\
0 & 0 & \alpha/2A & 0 & 0
\end{array} \right) \; ,
\label{eq:matrixADM2}
\end{equation}
and source terms
\begin{eqnarray}
q_{D_\alpha} &=& - \alpha \left( f + \alpha f' \right) D_\alpha K \; , \\
q_{D} &=& - 2 \alpha D_\alpha \left( K - 4 K_B \right) \; , \\  
q_{D_B} &=& - 2 \alpha D_\alpha K_B \; , \\
q_{K} &=& - \frac{\alpha}{A} \left[ D_\alpha \left (D_\alpha
- \frac{D}{2} \right) - D_B \left (D + \frac{D_B}{2} \right)
- A K^2 \right. \nonumber \\
&+& \left. \frac{2}{r} \left( D_\alpha -D + D_B \right)
- \frac{2}{r^2 B} \left( A - B \right) \right] \; , \\
q_{K_B} &=& - \frac{\alpha}{2 A} \left[ D_B \left( D_\alpha 
- \frac{D}{2} \right) - 2 A K K_B \right. \nonumber \\
&+& \left. \frac{1}{r} \left( 2D_\alpha - D + 2D_B \right)
- \frac{2}{r^2 B} \left( A - B \right) \right] \; .
\end{eqnarray}
Finally, the Hamiltonian and momentum constraints take the form
\begin{eqnarray}
C_h &:=& - \partial_r D_B
+ \frac{D_B}{2} \left( D + \frac{D_B}{2} \right) \nonumber \\
&+& A K_B \left( 2 K - 3 K_B \right) \nonumber \\
&+& \frac{D - D_B}{r} + \frac{A - B}{r^2 B} = 0 \; ,
\label{eq:HamCon2} \\
C_m &:=& - \partial_r K_B + \left( \frac{D_B}{2} + \frac{1}{r}
\right) \left( K - 3 K_B \right) = 0 \; . \hspace{10mm}
\label{eq:MomCon2}
\end{eqnarray}


\subsubsection{Modifying the equations by using the
constraints}

It turns out that, as they stand, neither the matrix $\bf{A}$ of the
original ADM system, (\ref{eq:matrixADM1}), nor the one of the
rewritten system, (\ref{eq:matrixADM2}), has a complete set of
eigenvectors for all $f>0$, so the systems of evolution equations are
not strongly hyperbolic.  Interestingly, strong hyperbolicity only
fails for $f=1$, which corresponds to harmonic slicing (this only
happens in spherical symmetry, in the full 3-dimensional case strong
hyperbolicity fails for ADM much more severely).  Since harmonic
slicing is such an important condition for both theoretical and
practical reasons, the systems described above are not very useful.

Let us concentrate on the second system of evolution equations. By
making use of the constraint equations, (\ref{eq:HamCon2}) and
(\ref{eq:MomCon2}), its principal part can be modified to construct a
strongly hyperbolic system for all $f>0$.  In particular, adding
multiples of the constraints will modify the third and fifth columns
of the matrix~(\ref{eq:matrixADM2}).  We will consider adjustments to
the evolution equations for the $v$'s of the form
\begin{eqnarray}
\partial_t v_i  &+& \sum_{j=1}^{(n+m)} A_{ij} \partial_r v_j \nonumber \\
&+& \alpha \left( h_i A^{{\cal H}_i} C_h
+ m_i A^{{\cal M}_i} C_m \right) = q_i \; .
\end{eqnarray}
Here the terms $\{h_i,m_i\}$ are allowed to depend on $f(\alpha)$,
such that for harmonic slicing these coefficients reduce to constants.
Furthermore, the exponents $\{{\cal H}_i,{\cal M}_i\}$ are fixed by
looking at the characteristic matrix and demanding that for $f=1$ its
entries have homogeneous powers of $A$.  Doing this we find
\begin{eqnarray}
{\cal H}_{D_\alpha} &=& {\cal H}_D = {\cal H}_{D_B} = - 1/2 \; , \\
{\cal H}_{K} &=& {\cal H}_{K_B} = - 1 \; , \\
{\cal M}_{D_\alpha} &=& {\cal M}_D = {\cal M}_{D_B} = 0 \; , \\
{\cal M}_{K} &=& {\cal M}_{K_B} = - 1/2 \; ,
\end{eqnarray}
and the characteristic matrix then takes the general form
\begin{equation}
\label{eq:matrixA}
{\bf A} = \left( \begin{array}{ccccc}
0 & 0 & \alpha h_{D\alpha} / A^{1/2} & \alpha f & \alpha m_{D\alpha}  \\
0 & 0 & \alpha h_D / A^{1/2} & 2 \alpha & \alpha (m_D-8) \\
0 & 0 & \alpha h_{D_B} / A^{1/2} & 0 & \alpha (2 + m_{D_B}) \\
\alpha/A & 0 & \alpha (2 + h_K) / A & 0 & \alpha m_K / A^{1/2} \\
0 & 0 & \alpha (1/2 + h_{K_B}) / A & 0 & \alpha m_{K_B} / A^{1/2}
\end{array} \right) .
\end{equation}
We now need to determine the coefficients $h_i$ and $m_i$ in order to
obtain a well behaved system of evolution equations.


\subsubsection{Integrability and hyperbolicity}

Since the $D$'s arise as spatial derivatives of the lapse and metric
components, their evolution equations are obtained by taking a time
derivative of their definition and then changing the order of the
partial derivatives.  If one later adds multiples of the hamiltonian
and momentum constraints to the evolution equations for the $D's$, one
finds that whenever these constraints are violated the $D$'s in fact
cease to be derivatives of metric functions.  One consequence of this
is the fact that in the sources of the evolution equations for the
eigenfields $w_i$, the coefficients $c_{iii}$ will no longer be
invariant under rescalings of the form $\tilde{w}_i = \Omega
(\alpha,A,B) w_i$ (the proof presented at the end of section
Sec.~\ref{sec:sourcecriteria} of the invariance of these coefficients
under such rescalings relied on the derivative constraints being
satisfied).  Such a property of our system of equations is
undesirable, as it makes the source criteria for avoiding blow-ups
impossible to apply in practice.

This leads us to the ``integrability criteria'', which states that the
$D$'s should remain derivatives of the metric functions independently
of the constraints, and implies that we must set the $h_D$'s and
$m_D$'s to zero and consider only adjustments in the evolution
equations of the extrinsic curvature variables.

Doing this we obtain for the characteristic matrix~(\ref{eq:matrixA})
the following eigenvalues
\begin{eqnarray}
\lambda_0 &=& 0 \; , \\
\lambda_\pm^f &=& \pm \alpha \sqrt{\frac{f}{A}} \; , \\
\lambda_\pm^c &=& 
\frac{\alpha}{2\sqrt{A}} \left( m_{K_B}
\pm \sqrt{4 + 8h_{K_B} + m_{K_B}^2} \right) \; .
\hspace{7mm}
\end{eqnarray}
The first three eigenvalues $(\lambda_0,\lambda^f_\pm)$ are precisely
the ones we found for the 1+1 dimensional case, so clearly $\lambda^f_\pm$ are
again gauge speeds (they depend on the gauge function $f$).  The last
two eigenvalues depend on the choices we have made to add constraints
to the evolution equations, so we will call them ``constraint
speeds''.  In fact, it is not surprising that we find only
characteristic speeds associated with the gauge and the constraints,
since in spherical symmetry it is well known that there are no
gravitational waves, i.e. no physical waves propagating at the speed
of light.

From the last expressions we see that if we want the constraint speeds
to be centered along the time lines we must ask for
\begin{equation}
m_{K_B} = 0 \; .
\end{equation}
If we now rewrite the parameter $h_{K_B}$ as
\begin{equation}
h_{K_B} = - \frac{1}{2} + \frac{\mu^2}{2} f \; ,
\label{eq:hKBmu}
\end{equation}
with $\mu$ a constant, then the eigenvalues
$\lambda^c_\pm$ take the following simple form
\begin{equation}
\lambda_{\pm}^{c} = \pm \mu \alpha \sqrt{\frac{f}{A}} \; .
\label{eq:mu_speed}
\end{equation}

At this point, only the adjustments to the evolution equation of the
trace of the extrinsic curvature, $h_K$ and $m_K$, and the constant
$\mu$ remain as free parameters.  It is now not difficult to show that
the system of evolution equations will be strongly hyperbolic, i.e.\
the matrix~(\ref{eq:matrixA}) can be diagonalized and a complete set
of eigenvectors exists, as long as $f>0$ and $\mu \notin \{ 0, \pm 1 \}$.
We want to point out here that the latter condition implies that all
characteristic speeds have to be distinct.  However, there is one
important exception, since for $|\mu| = 1$ the adjustments $h_K = -2$
together with $m_K = 0$ also yield a strongly hyperbolic system (which
will be of some importance later).  Furthermore we observe that no
particular problems arise for the case of harmonic slicing $(f = 1)$.
For completeness we explicitly state here the general form of the
eigenfields:
\begin{eqnarray}
w_{0} &=& D_\alpha - \frac{f}{2} D - 2 f D_B \;, 
\label{eq:symw0} \\
w_{\pm}^f &=& \sqrt{f} \left( 1 - \mu^2 \right) \left[ D_\alpha 
\pm \sqrt{f A} \; K \right] \nonumber \\
&+& \left[ \sqrt{f} \left( 2 + h_K \right)
\pm \frac{m_K f \mu^2}{2} \right] D_B \nonumber \\
&\pm& \left[ 2 \left( 2 + h_K \right)
\pm m_K \sqrt{f} \right] \sqrt{A} K_B \; , 
\label{eq:symwfpm} \\
w_{\pm}^c &=& \mu \sqrt{f} \; D_B \pm 2 \sqrt{A} \; K_B \; .
\label{eq:symwcpm}
\end{eqnarray}

In the case when $|\mu| = 1$, $h_K = -2$ and \mbox{$m_K = 0$}, the
eigenfields $w_{\pm}^f$ and $w_{\pm}^c$ take a different form given by
\begin{eqnarray}
w_{\pm}^f &=& \sqrt{fA} \; K \pm D_\alpha \; , 
\label{eq:specialwf} \\
w_{\pm}^c &=& \sqrt{f} \; D_B \pm 2 \sqrt{A} \; K_B \; .
\label{eq:specialwc}
\end{eqnarray}


\subsubsection{Indirect Linear Degeneracy}

Just as before, the linear degeneracy criteria is trivially satisfied.
Applying the indirect linear degeneracy criteria, $\partial
\dot{\lambda}_i / \partial w_i = 0$, to the eigenvalues $\lambda^f_\pm$,
we find 
\begin{equation}
\frac{\partial \dot{\lambda}_{\pm}^f}{\partial w_{\pm}^f}
= \frac{\alpha^2}{2 \left( 1 - \mu^2 \right) \sqrt{f} A} 
\left( 1 - f - \frac{\alpha f'}{2} \right) \; .
\label{eq:iLDgauge}
\end{equation}
This gives us the same condition on $f$ for avoiding a blow-up as
before, namely condition~(\ref{eq:gaugeshocks}), which is precisely
the result we obtained for the 1+1 dimensional case.  In addition,
however, we note that blow-ups can also arise from the second pair of
eigenvalues, for which we find
\begin{eqnarray}
\frac{\partial \dot{\lambda}_{\pm}^c}{\partial w_{\pm}^c}  
& = & - \frac{\mu \alpha^2}{4 \sqrt{f} A}
\left[ \left( 1 - f - \frac{\alpha f'}{2} \right) \right. \nonumber \\
& \times & 
\left. \left( \frac{4 + 2 h_K \pm \mu m_K \sqrt{f}}{1 - \mu^2} \right) 
+ 2f \right] \; . 
\label{eq:iLDconstraint}
\end{eqnarray}

The condition for avoiding these blow-ups is then
\begin{equation}
\mu \left[
\left( 1 - f - \frac{\alpha f'}{2} \right)
\left( \frac{4 + 2 h_K \pm \mu m_K \sqrt{f}}{1 - \mu^2} \right) 
+ 2 f \right] = 0 \; . 
\end{equation}
The first thing to notice is that we clearly must take 
\begin{equation}
m_K=0 \; ,
\end{equation} 
in which case the condition reduces to
\begin{equation}
\mu \left[
\left( 1 - f - \frac{\alpha f'}{2} \right)
\left( \frac{4 + 2 h_K}{1 - \mu^2} \right) 
+ 2 f \right] = 0 \; .
\label{eq:ilDconstraintshock}
\end{equation}
Now, if we insert a member of the gauge shock avoiding family
into this condition, we find
\begin{equation}
\mu f = 0 \; ,
\end{equation}
which, remembering that for strong hyperbolicity one must have $f > 0$
and $\mu \neq 0$, brings us to the rather discouraging result that - for the 
adjustments considered here - we can not avoid both gauge shocks and 
constraint shocks coming from the indirect linear degeneracy criteria at the 
same time.

Instead of using a full blown member of the gauge shock avoiding
family, we could be less ambitious and use a solution that avoids
gauge shocks only approximately.  For example, in
Ref.~\cite{Alcubierre02b} it was shown that the standard 1+log
slicing, corresponding to the choice $f=2/\alpha$, avoids gauge shocks
to first order (which explains why it is so robust in practice).
Taking such a form of $f$ we find that the condition for avoiding
constraint shocks is
\begin{equation}
h_K = - 2 \; \frac{\alpha - \mu^2}{\alpha - 1} \; .
\label{eq:ilD_1+log}
\end{equation}
For $|\mu| \approx 1$ this condition implies $h_K \approx -2$ which,
as we have seen in the previous section, is the only value of $h_K$ that 
yields a strongly hyperbolic system when $|\mu|=1$.  However, as mentioned
before, for the special case $\{|\mu|=1,h_K=-2\}$ the eigenfields are
different and the analysis should be performed separately.  When we
apply the indirect linear degeneracy criteria to the eigenfields
$w^c_\pm$ given by (\ref{eq:specialwc}), we find that $\partial
\dot{\lambda}^f_\pm / \partial w^f_\pm$ is non-zero, and since in this
case we have no free parameters, it is hard to say if the condition is
``closely satisfied'' or not.  Still, Eq.~(\ref{eq:ilD_1+log}) does
seem to indicate that the combination $\{|\mu|=1,h_K=-2\}$ is
preferred by indirect linear degeneracy. We will return to this case
when we consider some numerical examples below.


\subsubsection{Source Criteria}

The source criteria for the gauge eigenfields $w^f_\pm$ yields the
same condition on $f$ as indirect linear degeneracy since the
quadratic coefficient turns out to be
\begin{equation}
c_{\pm\pm\pm}^{fff} = \pm \frac{\alpha}{2 \left( 1 - \mu^2 \right) f \sqrt{A}}
\left( 1 - f - \frac{\alpha f'}{2} \right) \; .
\label{eq:sourcegauge}
\end{equation}
On the other hand, for the constraint eigenfields $w^c_\pm$ we find
the following quadratic coefficient
\begin{eqnarray}
c_{\pm\pm\pm}^{ccc} &=& \pm \frac{(1+\mu^2 f)\alpha}{16\mu^2 f \sqrt{A}}
\nonumber \\
&\times& \left[ 7 + 4h_K - 3 \mu^2 f \pm 2\mu m_K \sqrt{f} \right] \; .
\label{eq:sourceconstraint}
\end{eqnarray}
If we want to avoid a blow-up this coefficient must vanish,
\begin{equation}
 7 + 4h_K - 3 \mu^2 f \pm 2\mu m_K \sqrt{f} = 0 \; .
\end{equation}
This can be accomplished for any $f$ if we choose
\begin{equation}
m_K = 0 \; ,
\label{eq:sourcemK}
\end{equation}
and
\begin{equation}
h_K = -2 + \frac{1 + 3 \, \mu^2 f}{4} \; .
\label{eq:sourcehK}
\end{equation}
We first notice that again we obtain the condition \mbox{$m_K=0$},
just as we found in the previous section. It is also interesting to
notice that for a given choice of $f$ we have a one parameter family
of solutions that avoids these type of shocks. From (\ref{eq:hKBmu})
and (\ref{eq:sourcehK}) we see that the parameter $\mu$ relates $h_K$
and $h_{K_B}$ according to
\begin{equation}
h_K = -1 + \frac{3 \, h_{K_B}}{2} \; .
\label{eq:hKhKBrelation}
\end{equation}
Considering the restrictions on $\mu$ imposed by strong hyperbolicity,
$\mu \notin \{ 0, \pm 1\}$, we see that in the $(h_K,h_{K_B})$ plane
this shock avoiding family must be such that
\begin{equation}
h_{K_B} > - \frac{1}{2} \; , \qquad
h_{K_B} \neq \frac{1}{2} (f-1) \; .
\end{equation}

We see hence that by appropriate choices of $f(\alpha)$ and suitable
adjustments to the evolution equations, it is possible to avoid at the
same time the gauge and constraint shocks identified by the source
criteria.

As a final comment it is important to point out that, in contrast to
what we found for the gauge shocks, in this case indirect linear
degeneracy and the source criteria yield different statements for
avoiding constraint shocks.


\subsection{Numerical tests in spherical symmetry}

We will now describe some numerical experiments designed to test the
shock avoiding conditions found in the previous sections.  We will
concentrate on two different types of tests: The robust stability
test~\cite{Alcubierre03d}, and a test of Minkowski initial data with a
violation of the constraints added by hand.


\subsubsection{Robust stability test}

As a first numerical experiment we have performed the robust stability
test as described in~\cite{Alcubierre03d}. For this test one takes
Minkowski initial data and adds random noise with a small amplitude to
all dynamical variables.  For the evolution we have used both harmonic
slicing with $f=1$ and standard 1+log slicing with $f=2/\alpha$.

For the simulations discussed below we used 1,000 grid points, a grid
spacing of $\Delta r = 0.001$ and a time step $\Delta t = \Delta r/2$.
Furthermore, we have demanded periodic boundary conditions and have
set all $1/r$ and $1/r^2$ terms to zero by hand (i.e. we are
performing the run ``at infinity'').  Setting these terms to zero
should have no important effect on the presence of blow-ups since one
can readily verify that such terms are only linear in the eigenfields.
Moreover, one could easily regularize the equations at the origin by
following the procedure described in~\cite{Alcubierre04a}, but doing
so would only obscure the study of blow-ups by mixing them with purely
geometric effects.

We start the simulations with Minkowski initial data plus random
noise of order $10^{-6}$ on all evolution variables.  At latter times
we compute the error $\delta$ as the average absolute value over the
grid of the quantity $(\sum_{i=1}^3 |u_i - 1| + \sum_{i=1}^5
|v_i|)/8$.

\begin{figure}
\epsfxsize=80mm
\epsfysize=115mm
\epsfbox{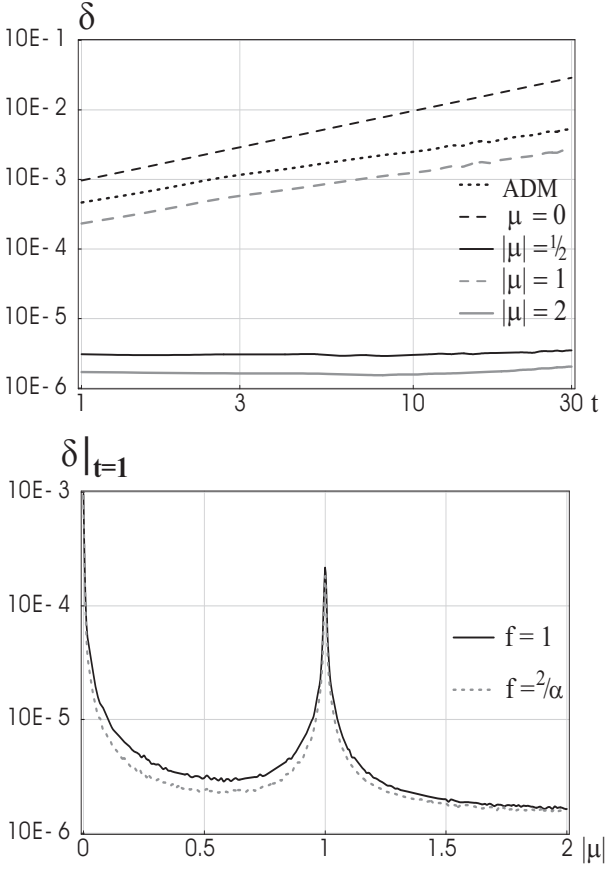}
\caption{\textit{Top.} Using harmonic slicing, the average error for
the robust stability test is shown for ADM and different members of
the $\mu$-family as a function of time (measured in units of the light
crossing time of our grid).  \textit{Bottom.} The value of the average
error after one light crossing time is plotted as a function of the
parameter $\mu$.}
\label{fig:mu}        
\end{figure}

We first performed runs for the ADM system without adding constraints,
and observed the well known growth in the average error caused by the
fact that ADM is only weakly hyperbolic.  Next, we implemented the
$\mu$-family given by (\ref{eq:hKBmu}) and (\ref{eq:sourcehK}), which 
according to the source criteria is shock-avoiding.  Figure~\ref{fig:mu} 
shows the behavior of the average error for this family.  In the top 
panel we plot the growth of the error for the case of harmonic slicing as 
a function of time (measured in units of the light-crossing time of our
computational grid), for several different values of $|\mu|$.  We see
that for ADM (which is not a member of the $\mu$-family), and for the
cases with $\mu=0$ and $|\mu| = 1$, the error grows linearly with
time, while for $|\mu| = 1/2$ and $|\mu| = 2$ the error initially does
not grow at all (at later times, however, also in these cases a linear
growth with a very small gradient develops). The lower panel in this
figure shows the value of the average error after one light-crossing
time as a function of $|\mu|$, as obtained both for harmonic and 1+log
slicings.  We see that for values of $|\mu|$ close to 0 or 1, the
error is already very large after one light crossing time, while away
from these values the error remains small.  The poor behavior of the
simulations with $\mu=0$ and $|\mu| = 1$ is caused by the fact that
for such cases the evolution system is not strongly hyperbolic.  Also,
from~(\ref{eq:mu_speed}) we see that for values of $|\mu|$ close to but
different from either $0$ or $1$, the eigenspeeds $\lambda^c_\pm$ 
associated with the constraint modes become very similar to either 
$\lambda_0$ or $\lambda^f_\pm$.  This means that even if the system is 
still strongly hyperbolic, the argument used for ignoring mixed terms 
in the sources will not apply.

\begin{figure}
\epsfxsize=80mm
\epsfysize=115mm
\epsfbox{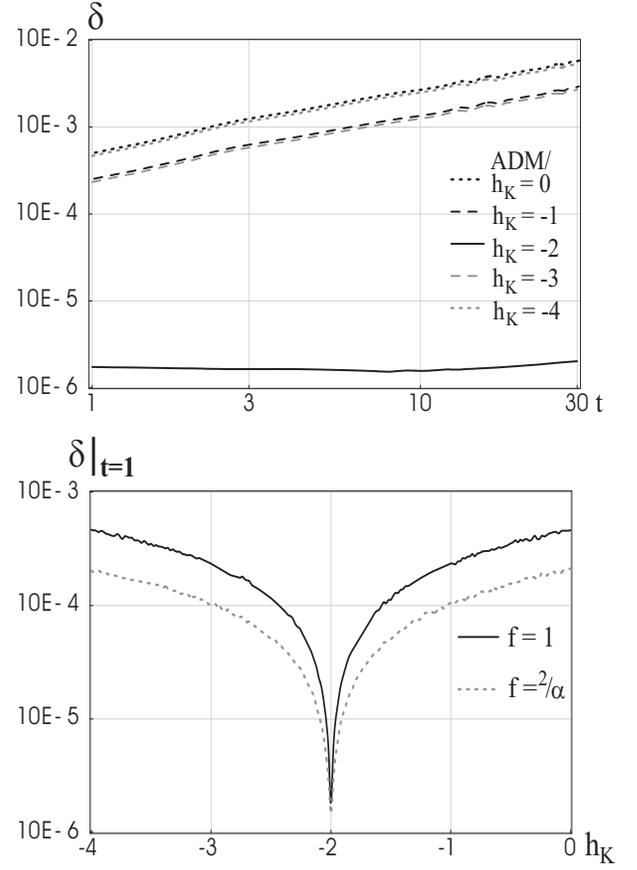}
\caption{\textit{Top:} Setting $|\mu| = 1$, the average error for
  harmonic slicing is found to grow only moderately for \mbox{$h_K =
    -2$}, whereas for other values of $h_K$ (and in particular for ADM
  corresponding to $h_K = 0$) a rapid growth is observed.
  \textit{Bottom:} The average error after one crossing time is
  plotted as a function of $h_K$.}
\label{fig:hK}        
\end{figure}

In particular, for the case $|\mu| = 1$ the adjustment \mbox{$h_K =
  -2$} suggested by the indirect linear degeneracy criteria turned out
to be helpful.  Because of this in Figure~\ref{fig:hK} we also show
similar plots testing different values of the parameter $h_K$ in the
case of $|\mu| = 1$.  Here one can observe that for values of $h_K$
other than -2, and in particular for ADM corresponding to $h_K = 0$, a
linear growth in the average error is present.  For the adjustment
suggested by the indirect linear degeneracy criteria (which is the only 
value here which yields a strongly hyperbolic system), initially no error
growth is found.

\subsubsection{Minkowski initial data plus constraint violation}
\label{sec:Minkwoski_constraint}

As we saw in the previous section, the robust stability test is very
good at distinguishing between strongly and weakly hyperbolic systems,
but does not show any clear indication that, among strongly hyperbolic
systems, those that avoid shocks are better behaved.  This should not
be surprising as the robust stability test uses random, uncorrelated
and non-smooth initial data, while shock formation is the result of
the coherent evolution of smooth initial data.

For this reason, using harmonic slicing we have performed evolutions
of Minkowski initial data with a smooth violation of the
constraints added by hand.  Here we have chosen a perturbation in
$K_B$ similar to the one we used for the scalar wave equation, namely
\begin{equation}
K_B (t=0) = - \left( 2 \kappa r /\sigma^2 \right) \,
\exp \left( -r^2/\sigma^2 \right) \; ,
\end{equation}
with parameters $\kappa = 0.05$ and $\sigma = 1$.

For the simulations shown below we used 8,000 grid points and a
grid spacing of $\Delta r = 0.01$ (which places the boundaries at $\pm
40$) together with a time step \mbox{$\Delta t = \Delta r/2$}.
Furthermore, we have again neglected $1/r$ and $1/r^2$ terms as the
simulation can be ``shifted'' to large values of $r$.

\begin{figure}
\epsfxsize=80mm
\epsfysize=125mm
\epsfbox{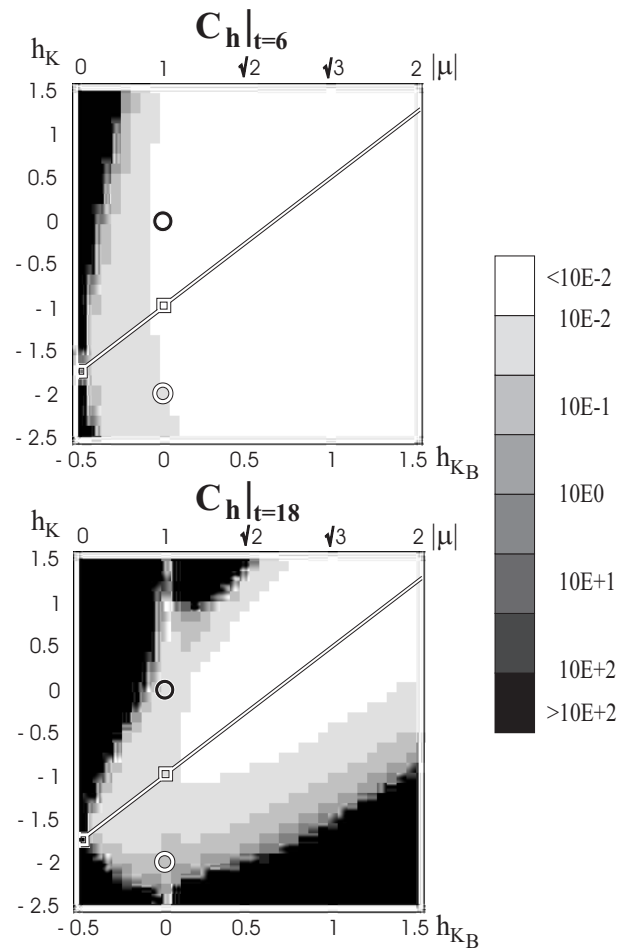}
\caption{Contour plots of the r.m.s. of the Hamiltonian constraint are
shown at times $t = 6$ and $t = 18$.}
\label{fig:contour}        
\end{figure}

In Figure~\ref{fig:contour} we show contour plots of the root mean
square (r.m.s.) of the Hamiltonian constraint as a function of the
adjustment parameters $h_K$ and $h_{K_B}$ at two different times
during the evolution, using 40 equidistant parameter choices in each
direction.  The momentum constraint, not shown here, has a very
similar behavior.  Note that black and dark gray colors correspond to
regions where the r.m.s. of the Hamiltonian constraint grows rapidly,
and light gray denotes regions where it grows very slowly or not at
all.  For ADM (corresponding to $h_K = h_{K_B} = 0$ and denoted by a
black circle) a rapid growth of the constraints is found. We also
observe rapid growth close to the white circle, representing the
special case with $h_K=-2$ and $h_{K_{B}}=0$ which corresponds to the 
only strongly hyperbolic system along the line $|\mu|=1$ and is 
preferred by indirect linear degeneracy.

The white diagonal line corresponds to the shock avoiding $\mu$-family
obtained from the source criteria (not defined for the points
\mbox{$\mu = 0$} and \mbox{$|\mu| = 1$} represented as boxes).  We
clearly see that this one-parameter family falls in the middle of the
region where the r.m.s. of the Hamiltonian constraint does not grow,
indicating that it does correspond to a preferred region of parameter
space.

There is a final point related to Figure~\ref{fig:contour}.  The
figure shows that the line $|\mu|=1$ also seems to produce slow growth
of the constraints.  However, as mentioned before, in that case the
system is only weakly hyperbolic and our whole analysis breaks down,
so we have no explanation as to why this line represents a preferred
region.

\begin{figure}
\epsfxsize=80mm
\epsfysize=150mm
\epsfbox{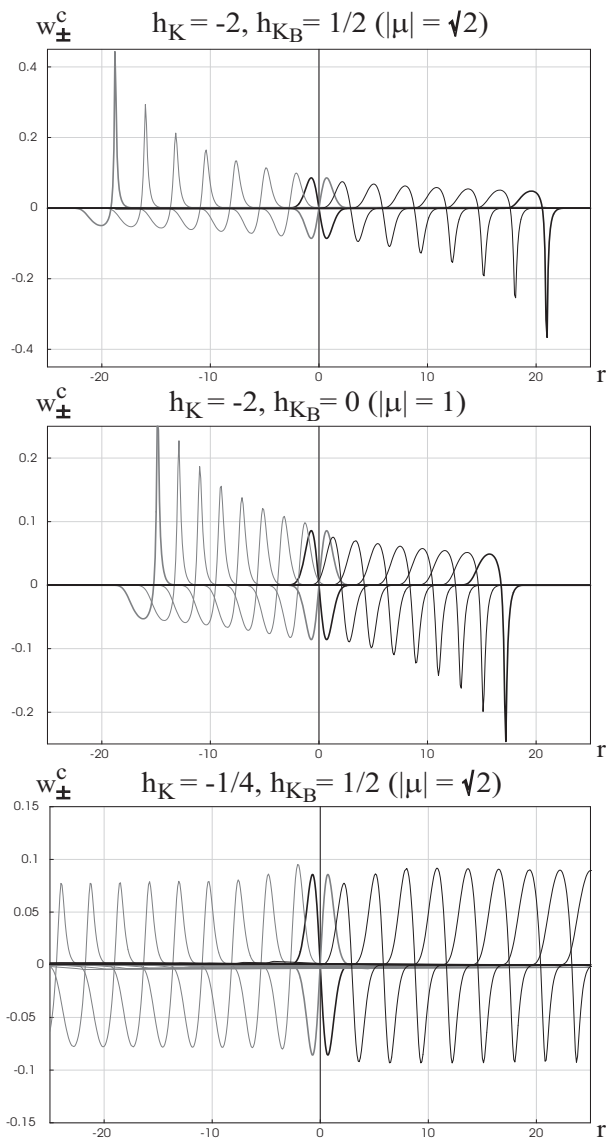}
\caption{Evolution of the eigenfields $w^c_+$ (black) and $w^c_-$
(gray) associated with constraint shocks.  \textit{Top.} Blow-ups
occur for a strongly hyperbolic case that is not shock avoiding.
\textit{Middle.} Shocks also form for the adjustments suggested by the
indirect linear degeneracy criteria.  \textit{Bottom.} No shocks are
found for a member of the $\mu$-family obtained by the source
criteria.}
\label{fig:KBws}        
\end{figure}

In order to see the formation of shocks more clearly, in
Figure~\ref{fig:KBws} we show the time evolution (shown every
\mbox{$\Delta t = 2$}) of the eigenfields $w^c_\pm$ which are
associated with the formation of constraint shocks.  The upper panel
shows the evolution for the parameter choice $h_K=-2$ and
$h_{K_B}=1/2$ (which for harmonic slicing implies $|\mu| = \sqrt{2}$),
corresponding to a strongly hyperbolic case that does not avoid
constraint shocks.  From the figure we see how a shock is clearly
forming, as expected.  The middle panel corresponds to the parameters
$h_K=-2$ and $h_{K_B}=0$ (and hence $|\mu| = 1$) preferred by the
indirect linear degeneracy criteria.  Again we observe the formation
of shocks.  Finally, the lower panel shows the evolution for a member
of the shock-avoiding $\mu$-family with parameters $h_K = -1/4$ and
$h_{K_B} = 1/2$ (i.e., $|\mu| = \sqrt{2}$).  In this case the
evolution shows a wave-like character, and no shocks form on the
time-scale considered here.

\section{Conclusions}
\label{sec:conclusions}

We have presented an analysis of two different blow-up mechanisms for systems 
of hyperbolic equations of the type found in general relativity, namely 
the geometric blow-up or gradient catastrophe, and the ODE-mechanism.  
As an example of how these mechanisms operate we have used the simple 
one-dimensional wave equation with dynamic wave speed and non-trivial 
source terms.

We have later performed a blow-up analysis of one-dimensional systems
in general relativity, concentrating on ``toy'' 1+1 relativity and
spherically symmetric relativity, and using the hyperbolic Bona-Masso
family of slicing conditions.  In the first case we have recovered the
well known gauge shocks and found, somewhat surprisingly, that both
blow-up criteria give precisely the same condition for shock
avoidance.  When studying the spherically symmetric case, we have also
found the same type of gauge shocks, but much more importantly, we
have found a second family of modes that can produce blow-ups.  The
characteristic speed of these extra modes is controlled by the way in
which the constraint equations are used to modify the evolution
equations, so we have called such blow-ups ``constraint shocks''.  We
have also studied how one can adjust the way in which constraints are
added to the evolution equations to avoid as much as possible the
formation of such constraint shocks.  Finally, we have presented
numerical simulations that confirm that constraint shocks can and do
indeed form, and that they can be avoided by using the results of our
previous analysis.  

We would like to mention here that although in the present study we did 
not work with adjustments that can avoid constraint shocks according to 
both blow-up criteria, this is simply because we have considered only a 
restricted form of adjustments to the evolution equations.  
More elaborate adjustments can give rise to completely shock free 
formulations~\footnote{In fact, it is not very difficult to find 
adjustments that satisfy both criteria.  By making use
of an arbitrary function $c(\alpha,A,B)$, the adjustment~(\ref{eq:hKBmu}) 
can by generalized to obtain $h_{K_B} = (c-1)/2$.  The source criteria 
then implies that \mbox{$h_K = -2 + (1 + 3 c)/4$}, which generalizes 
the previous expression~(\ref{eq:sourcehK}) and again yields the 
relation~(\ref{eq:hKhKBrelation}).  Furthermore, since the
eigenvalues $\lambda^c_\pm$ are now given by \mbox{$\lambda^c_\pm =
\pm \alpha \sqrt{c/A}$}, by choosing $c = \mu^2 A/\alpha^2$ one obtains
$\lambda^c_\pm = \pm \mu = {\rm const}$, which trivially satisfies the
indirect linear degeneracy criteria.  Numerical experiments show that
systems with these adjustments behave in a very similar way to those 
described in the text.}.  Moreover, empirically we have found that 
the source criteria seems to be far more important than indirect linear
degeneracy (though it is still not clear why in the case of gauge
shocks both criteria give rise to the same condition for avoiding
blow-ups).

As a final note we want to point out that there is a very important
difference between gauge and constraint shocks.  When a gauge shock
develops, it means that our coordinate system has broken down and
there is no way to continue the evolution past that point other than
choosing a different gauge. Gauge shocks are certainly non-physical,
as the geometry of spacetime remains smooth, but they do represent
very real pathologies in the spatial foliation.  Constraint shocks, on
the other hand, are not only non-physical, but they should not arise
at all if the constraints remain satisfied exactly. In a numerical
simulation the constraints are of course violated, but as the
resolution is increased this violation should become smaller and
smaller, so the possible formation of constraint shocks becomes less
of an issue.  Nevertheless, at any fixed resolution constraint shocks
can form at a finite time unless care is taken to use a form of the
equations that avoids them.

Because of this we recommend that out of the many possible ways of
constructing strongly hyperbolic evolution systems in general
relativity, for numerical evolutions one should concentrate on those
that have better shock avoiding properties, both in the gauge and
constraint violating sectors, as these should prove to be more robust
in practice.


\acknowledgments

It is a pleasure to thank Marcelo Salgado, Alejandro Corichi, Denis
Pollney, Sascha Husa and Olivier Sarbach for many useful discussions
and comments.  We also want to thank an anonymous referee for many
detailed and useful comments that have helped to improve significantly
our original manuscript. This work was supported in part by DGAPA-UNAM
through grants IN112401 and IN122002, and by NSF grant PHY-02-18750.
B.R.  acknowledges financial support from the DAAD and the
Richard-Schieber Stiftung.


\bibliographystyle{bibtex/apsrev}
\bibliography{bibtex/referencias}


\end{document}